\documentclass{article}

\usepackage{arxiv}

\usepackage[utf8]{inputenc} 
\usepackage[T1]{fontenc}    
\usepackage{hyperref}       
\usepackage{url}            
\usepackage{booktabs}       
\usepackage{amsfonts}       
\usepackage{nicefrac}       
\usepackage{microtype}      
\usepackage{lipsum}
\usepackage{graphicx}
\graphicspath{ {./images/} }

\usepackage{dirtytalk}
\usepackage{subcaption}
\usepackage{amsmath}

\title{Excited-State Intramolecular Proton Transfer and Competing Pathways in 3-Hydroxychromone: A Non-adiabatic Dynamics Study}

\author{
 Alessandro Nicola Nardi \\
  Nantes Universit\'{e}, CNRS, CEISAM, UMR 6230\\
  Nantes F-44000, France \\
   \And
 Morgane Vacher$^\ast$ \\
  Nantes Universit\'{e}, CNRS, CEISAM, UMR 6230\\
  Nantes F-44000, France \\
  \texttt{morgane.vacher@univ-nantes.fr} \\
}

\begin{document}
\maketitle
\begin{abstract}
Excited-state intramolecular proton transfer (ESIPT) is a fundamental photochemical process in which photoexcitation induces proton transfer within a molecule, leading to the formation of a tautomeric excited state.
It was observed experimentally that the 3-hydroxychromone (3-HC) system exhibits two distinct proton-transfer time scales upon excitation to the lowest \say{bright} singlet excited state: an ultrafast component on the femtosecond time scale and a slower one on the picosecond time scale, largely insensitive to solvent effects.
Up to now, the microscopic origin of the second time constant has only been hypothesised.
Here, using mixed quantum-classical non-adiabatic dynamics simulations, we explicitly observe the two ESIPT time constants and we rationalise the origin of the second time scale by the presence of a competitive out-of-plane hydrogen torsional motion.
Comprehensive analysis of the excited-state potential energy surfaces and nonadiabatic trajectories enables us to construct an explicit reaction network for 3-HC, delineating the interplay between canonical ESIPT and torsion-mediated pathways. This unified mechanistic framework reconciles the coexistence of ultrafast and slower ESIPT components, offering new insights into the non-adiabatic excited-state dynamics of the system.
\end{abstract}


\section{Introduction}
Excited-state intramolecular proton transfer (ESIPT) refers to a photochemical event where photoexcitation triggers proton transfer within a molecule, producing a tautomeric excited state.
The earliest description of this process was provided by Weller \cite{weller1955fluoreszenz,weller1956innermolekularer}, noting that the remarkable Stokes shift (of nearly 10,000 cm$^{-1}$) detected in the fluorescence of salicylic acid could be explained by a fast excited-state proton transfer. Indeed, the ESIPT induces a major structural reorganisation compatible with the large observed shift.
Later, Ottersted \cite{otterstedt1973photostability} and Kasha \cite{sengupta1979excited} also focused on the interpretation of this phenomenon in other molecular systems.

Since then, significant experimental and theoretical computational efforts have been dedicated to understanding excitation-induced proton transfer in a wide range of molecular systems. This resulted in numerous applications of ESIPT systems as fluorescent probes \cite{zhao2012excited,sedgwick2018excited,li2021progress}, photostabilizers \cite{paterson2005mechanism,gong2016new}, organic light-emitting diodes (OLEDs) \cite{kwon2011advanced,tang2011fine,zhang2016control,mamada2017highly,li2018dual,wu2018novo}, and in recognition of its central role in several biochemical processes \cite{sytina2008conformational,tonge2009excited,weinberg2012proton,jacquemin2014assessing,joshi2021excited}.
The time scale of the ESIPT process is considered rather short \cite{zhou2018unraveling}, on the femtosecond to picosecond time scale, depending on the molecular properties that can also affect the yield.
In particular, the presence of an activation barrier along the coordinate that connects the relevant tautomers involved in the proton transfer
and/or the competition with other molecular motions can lead to a notable reduction of the rate and yield in comparison to the case in which the ground-state tautomer is not stable in the excited-state and transforms, in a ballistic fashion, toward the photoproduct. In the case of incomplete tautomerization, the emission spectrum can exhibit two bands: one with a modest Stokes shift from the reactant state and another with a large Stokes shift from the ESIPT product. The latter aspect is also referred to as dual fluorescence.

Understanding the dynamics and mechanisms of such a process is important not only from a fundamental photochemical and photobiological perspective but also crucial for the rational design of molecules with tailored optical properties.
The experimental investigation of the ultrafast ESIPT requires high-resolution time-resolved spectroscopic techniques. Among them, femtosecond stimulated Raman spectroscopy (FSRS) offers valuable insight into skeletal motions during proton transfer \cite{fang2009mapping,han2013excited}, as well as time-resolved fluorescence, UV/vis and IR absorption spectroscopy \cite{chevalier2013esipt}.

Since it is difficult from experiments to obtain information about transition state structures, energy barriers,
possible conical intersections between the involved excited states, or other critical features of the reaction pathway, such data, essential for unraveling the detailed mechanism of ESIPT, are often obtained from state-of-the-art \textit{ab initio} excited-state calculations. In recent years, theoretical investigations have therefore been carried out on a wide variety of ESIPT systems using diverse computational strategies:
among them, time dependent-density functional theory (TD-DFT) \cite{laurent2014dye,houari2014modeling,chrayteh2020dual}, post-Hartree-Fock approaches, such as second order approximated coupled cluster (CC2) \cite{aquino2005excited,aquino2009ultrafast,sobolewski2009computational} or algebraic diagrammatic construction approximated to the second order (ADC(2)) \cite{verite2019theoretical,chrayteh2020td}, and multireference methods such as complete active space self-consistent field (CASSCF), in some studies with the inclusion of the perturbative correction (CASPT2) \cite{sobolewski1999ab,li2015casscf,chang2023caspt2}.
Along with static approaches, non-adiabatic dynamics simulations, employing multi-configuration time-dependent Hartree (MCTDH) \cite{ortiz2007electronic,perveaux2017fast,anand2020excited,anand2020h}, ab initio multiple spawning (AIMS) \cite{pijeau2017excited,pijeau2018effect}, and trajectory surface hopping (TSH) \cite{sporkel2013photodynamics,nag2021ultrafast,nag2022unraveling,zhao2024excited}, were performed to explicitly follow the proton transfer on the excited state surface(s).

\begin{figure}
     \centering
     \begin{subfigure}[c]{\linewidth}
         \centering
         \includegraphics[width=3.25in]{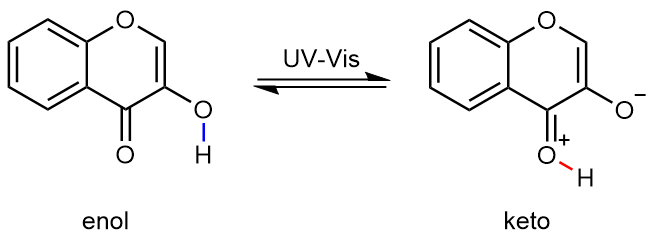}
         \caption{}
         \label{fig:3-HC_tautomers}
     \end{subfigure}
     \vfill
     \begin{subfigure}[c]{\linewidth}
         \centering
         \includegraphics[width=3.25in]{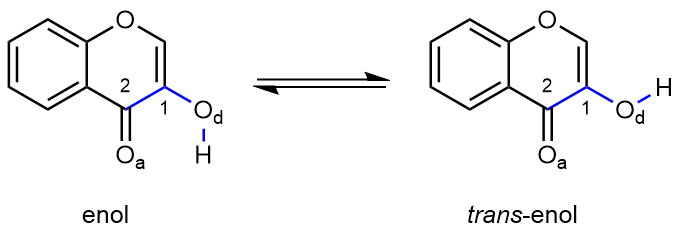}
         \caption{}
         \label{fig:3-HC_conformers}
     \end{subfigure}
        \caption{(a) Enol and keto tautomeric forms of the 3-hydroxychromone (3-HC). (b) (\textit{cis}-)enol to \textit{trans}-enol conformational equilibrium.}
        \label{fig:3-HC_tautomers_conformers}
\end{figure}

In recent years, 3-hydroxychromone (3-HC), depicted in Figure \ref{fig:3-HC_tautomers}, and its derivatives have drawn significant attention for their photochemical properties and potential applications as molecular probes that exploit their spectrally distinct dual fluorescence \cite{chevalier2013esipt}.
Indeed, upon excitation to the first (\say{bright}, $\pi\pi^*$) singlet excited-state, the 3-HC molecule can easily access the S$_1$ minimum and fluoresce or undergo fast ESIPT reaction. 
Chevalier et al. \cite{chevalier2013esipt} reported a distinctive behaviour of the 3-HC molecule, namely the presence of two ESIPT rate constants: a fast component on the femtosecond time scale and a slower one on the picosecond time scale, independent of the solvent environment. 
In polar protic solvents, such as water, the formation of the conjugated base of the molecule (i.e., 3-HC deprotonated at the hydroxyl functional group) and the formation of solute-solvent intermolecular hydrogen bonds could be competitive with the ESIPT process, and was thus hypothesised to give rise to the slower component in the picosecond time scale. 
Furthermore, in aprotic apolar solvents such as methylcyclohexane, a slower ESIPT component was also observed, and attributed either to (i) intramolecular vibrational relaxation (IVR), which reduces the excitation of proton transfer-promoting modes, and/or (ii) transient access to the excited \textit{trans}-enol conformation (see Figure \ref{fig:3-HC_conformers}).

Among theoretical-computational studies on the 3-HC system \cite{estiu1999theoretical,ash2011excited,kenfack2012ab,nag2021ultrafast,zhao2024excited} only Perveaux et al. \cite{perveaux2017fast} specifically addressed the origin of the slower (picosecond time scale) ESIPT rate constant. They localised a previously undocumented conical intersection between the first two singlet excited states (the \say{bright}, $\pi\pi^*$ and \say{dark}, $n\pi^*$, at the FC point) in 3-HC. This intersection exhibits a planar geometry (C$_s$ symmetry) and is readily accessible from the FC point. Using multilayer MCTDH quantum dynamics simulations, the authors proposed that following excitation to S$_1$, the 3-HC nuclear wave packet can reach this S$_1$/S$_2$ conical intersection, enabling ultrafast population transfer to the $n\pi^*$ state along the non-reactive direction of the ESIPT coordinate. These findings were later qualitatively supported by the quantum dynamics simulations of Anand et al. \cite{anand2020excited}.
However, the theoretical reproduction of two ESIPT time constants from non-adiabatic dynamics simulations and an explicit correlation with the hypothesised mechanism is still lacking. 

We present here a theoretical-computational investigation of the ESIPT reaction in the 3-HC molecule (Figure \ref{fig:3-HC_tautomers}) by means of on-the-fly mixed quantum-classical nonadiabatic dynamics simulations. Our aim is to contribute to a thorough mechanistic explanation of the factors that lead to the observed double time constants associated with the ESIPT reaction.
A deeper understanding of the mechanisms competing with the proton transfer reaction could aid in the design of more efficient 3-HC-based molecular probes.

The article is structured as follows: in Section \ref{sec:computational_details} the computational details used in this work are presented. In Section \ref{sec:results_discussion}, the results are exposed and discussed. Section \ref{sec:conclusions} offers some concluding remarks.

\section{Theoretical methods and computational details}
\label{sec:computational_details}

\subsection{Electronic structure methods}
At the Franck-Condon (FC) point, the 3-HC molecule exhibits two low-lying excited states that are nearly degenerate. Small geometric displacements from the FC point can readily alter the state character owing to this near-degeneracy of S$_1$ and S$_2$ states.
This emphasizes the necessity of (i) a rigorous characterization of the energetics and state characters involved, and (ii) the explicit inclusion of non-adiabatic effects in describing nuclear dynamics, even in the vicinity of the FC point.
Electronic structure calculations by Perveaux et al. \cite{perveaux2017fast} examined the reliability of TD-PBE0 in capturing the S$_1$ (\say{bright}, $\pi\pi^*$) and S$_2$ (\say{dark}, $n\pi^*$) states at the FC point and critical regions of the potential energy surfaces, through comparisons with the higher-level CC2, ADC(2), and EOM-CCSD methods.
Anand et al. \cite{anand2020excited} assessed the TD-DFT approach, reporting an S$_1$–S$_2$ energy gap at the FC point between 0.01 and 0.10 eV, depending on the functional, in qualitative agreement with the 0.15 eV gap obtained from their EOM-CCSD calculations.

On this basis, TD-PBE0 \cite{adamo1999toward} was chosen here to compute the potential energy surfaces and their couplings for the surface-hopping simulations. This functional offers a favourable balance between accuracy and computational cost and has been shown to reliably describe the electronic structure at key points along the ESIPT coordinate \cite{perveaux2017fast}. The calculations employed the cc-pVDZ \cite{dunning1989gaussian} basis set.
To asses the influence of the basis set, the critical points along the ESIPT pathway were re-optimised using the cc-pVTZ and aug-cc-pVDZ bases and the energy profiles were re-computed.
The S$_1$–S$_2$ energy gap at the FC point from our TD-PBE0/cc-pVDZ (cc-pVTZ/aug-cc-pVDZ) calculations is 0.12 (0.08/0.10) eV.
The results obtained from the non-adiabatic dynamics simulations will be compared (see below) with experimental data obtained in methylcyclohexane. Therefore, the influence of the solvent was tested on the aforementioned ESIPT profile using the dielectric continuum description provided by the IEFPCM \cite{lipparini2010variational}.
All the calculations were carried out with Gaussian 16, rev. A03 \cite{g16}. The tight SCF convergence criteria and the fine integration grid were requested.
Several key conical intersections were optimised with the Orca 6.1 package \cite{neese2025software} using the Updated Branching Plane (UBP) method, which circumvents the need for explicit non-adiabatic coupling matrix elements \cite{maeda2010updated}. The optimisation calculations employed the same level of theory as above, namely TD-DFT with the PBE0 functional and the cc-pVDZ basis set.

\subsection{Propagation of the coupled electron-nuclear dynamics}
Trajectory surface-hopping simulations were carried out with the SHARC software package (version 3.0.1) \cite{mai2018nonadiabatic}. Initial conditions were generated from the 0 K Wigner distribution based on the ground-state vibrational frequencies of the (\textit{cis}-)enol conformer. A total of 10,000 geometries were sampled, and for each, the excitation energies and oscillator strengths of the first two singlet states (S$_1$ and S$_2$) were computed at the TD-PBE0/cc-pVDZ level via the SHARC-GAUSSIAN interface. These data were employed to construct the absorption spectrum using the nuclear ensemble approach \cite{crespo2012spectrum}, where each line was convoluted with a Gaussian function of 0.1 eV full width at half maximum. No energy shift was applied to the computed spectrum with respect to the experimental one.

From this ensemble, the initial conditions for the non-adiabatic dynamics simulations were selected based on excitation energy and oscillator strength. Specifically, an energy window of 0.2 eV centered at 3.9 eV (ca. 320 nm, the excitation wavelength used by Chevalier et al. \cite{chevalier2013esipt} in their time-resolved UV–vis experiments) was applied. Within this range, the brightest geometries for each state were chosen following the stochastic selection procedure of Pite\v{s}a et al. \cite{pitesa2024excitonic}, as implemented in SHARC. This yielded 311 and 86 trajectories initiated on the S$_1$ and S$_2$ states, respectively.

A nuclear time step of 0.5 fs was used. Three electronic states, S$_0$-S$_2$, were considered in the non-adiabatic dynamics simulations. The electronic wave function was propagated using the local diabatization method \cite{granucci2001direct} with 25 substeps. Granucci and Persico decoherence scheme was used, with a factor of 0.1 a.u. of energy \cite{granucci2007critical}.
Hopping probabilities between S$_1$ and S$_2$ were computed from the time evolution of the electronic amplitudes. An energy threshold of 0.10 eV was used to enforce a hop between an active excited state and the ground state, due to the well-documented deficiency of TD-DFT in the description of conical intersections involving the electronic ground state \cite{levine2006conical}. After a hop, the nuclear velocity vector was rescaled isotropically to conserve the total energy.
All trajectories exhibit total energy conservation within the selected 0.5 eV threshold in both S$_1$ and S$_2$ ensembles. No trajectories terminated prematurely due to convergence failures in the electronic structure calculations.

\subsection{Analysis}
To gain deeper insight into the ESIPT process and competing relaxation pathways, structural and statistical analyses on the full ensemble of trajectories were performed. A chemically intuitive and practical way to characterize the ESIPT reaction is through the distances between the transferring proton and the two oxygen atoms involved. We denote these bond distances as $d(\mathrm{O_d}-\mathrm{H})$ and $d(\mathrm{H}-\mathrm{O_a})$, corresponding to the donor oxygen–hydrogen and hydrogen–acceptor oxygen distances, respectively (see Figure \ref{fig:3-HC_conformers}).
To probe the role of torsional motions in both tautomers, we additionally monitored the dihedral angles $\phi_1$ and $\phi_2$, defined by the atoms (H, $\mathrm{O_d}$, $\mathrm{C_1}$, $\mathrm{C_2}$) and (H, $\mathrm{O_a}$, $\mathrm{C_2}$, $\mathrm{C_1}$), respectively (see Figure \ref{fig:3-HC_conformers}).
All structural analyses were performed with the aid of the MDTraj software package \cite{McGibbon2015MDTraj}.

Along the mixed quantum–classical trajectories, the molecule was classified as either enol or keto (see Figure \ref{fig:3-HC_tautomers}) tautomer to obtain the time-dependent yield of the ESIPT reaction (see below). The (\textit{cis}-)enol form is the most stable tautomer in the ground state and is the form from which the dynamics starts. In contrast, the keto tautomer is stabilized in the S$_1$ state and is formed along the ESIPT reaction coordinate. The time-dependent yield of the keto form was monitored by analysing the distances $d(\mathrm{O_d}-\mathrm{H})$ and $d(\mathrm{H}-\mathrm{O_a})$ along the trajectories. At each time step, the structure was assigned to the keto form when $d(\mathrm{H}-\mathrm{O_a}) < d(\mathrm{O_d}-\mathrm{H})$.
To further analyse the nuclear motions along the non-adiabatic trajectories, a principal component analysis (PCA) \cite{pearson1901liii} was performed on the full ensemble of structures. This approach allowed us to identify the dominant collective motions contributing to the ESIPT process and other relevant conformational changes.

\section{Results and discussion}
\label{sec:results_discussion}
The results are now presented, starting with a description of the two lowest-lying valence excited electronic states at the ground state equilibrium geometry of 3-HC, its UV-vis absorption spectrum, and some critical points on the excited state surfaces along the ESIPT coordinate and possible competitive pathways.
Then, the photodynamics of 3-HC, as described by surface hopping simulations, is discussed. In particular, we will focus on the analyses of the adiabatic electronic state populations and the key nuclear degrees of freedom involved in the ESIPT reaction, the kinetic modelling of the proton transfer, and the definition of a complete reaction network that connects all the relevant species involved in their different electronic states.

\subsection{Vertical absorption spectrum}
\begin{figure}
    \centering
    \includegraphics[width=3.25in]{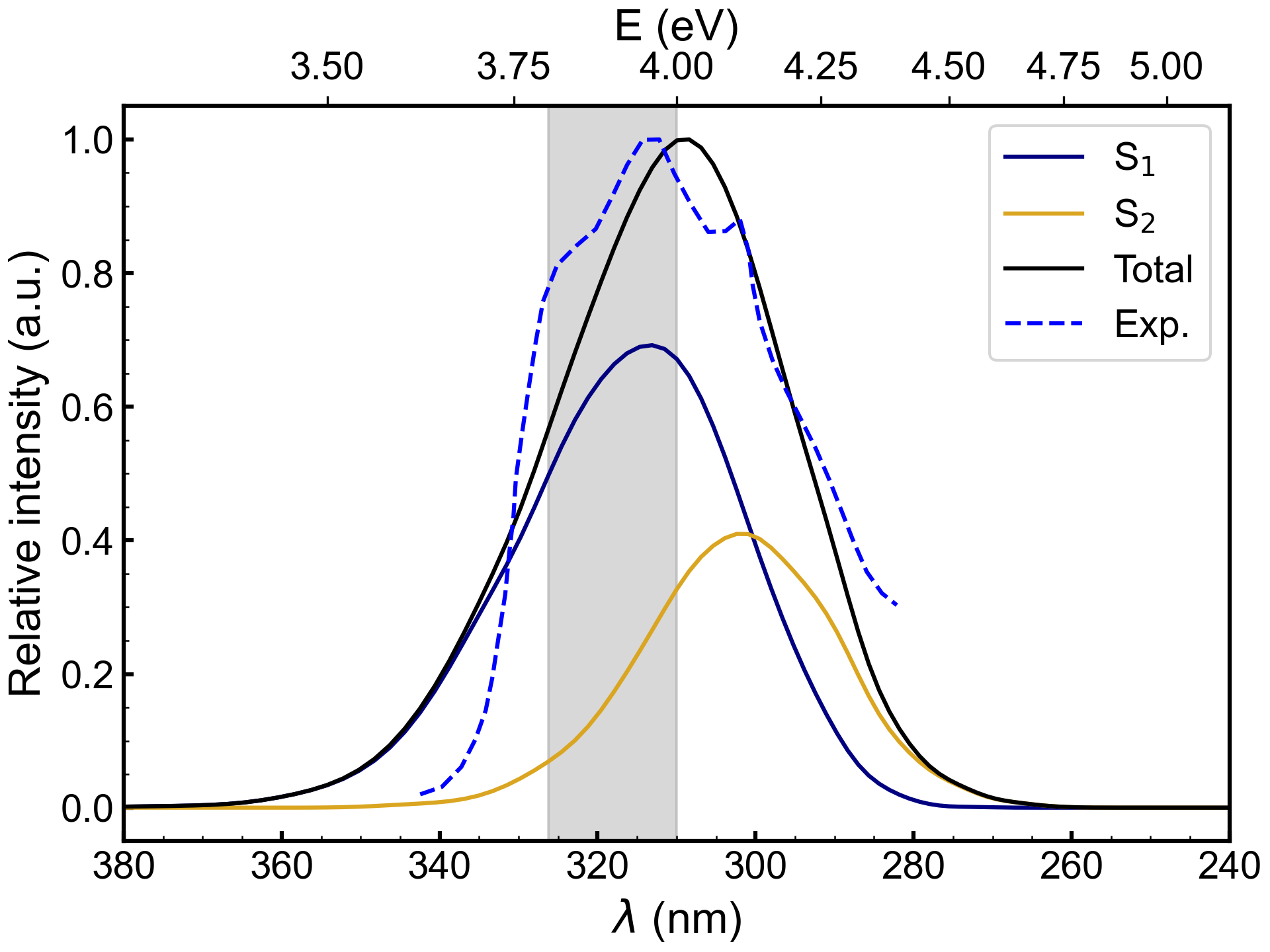}
    \caption{TD-PBE0/cc-pVDZ vertical absorption spectrum of 3-HC obtained via the nuclear ensemble approach. The gray band represents the energy window used for the initial condition selection. No energy shift was applied to the computed spectrum. The digitalised experimental absorption spectrum (band C) from Chevalier et al. \cite{chevalier2013esipt} is reported as a blue dashed line.}
    \label{fig:vert_spectrum_10000initconds}
\end{figure}

The vertical absorption spectrum of 3-HC in the gas phase was computed via the nuclear ensemble approach \cite{crespo2012spectrum} from all (10,000) geometries sampled from the Wigner distribution. Vertical excitation energies at each geometry were evaluated at TD-PBE0/cc-pVDZ level of theory, and a Gaussian line shape with FWHM of 0.1 eV was applied. The resulting spectrum is shown in Figure \ref{fig:vert_spectrum_10000initconds}. The computed maximum absorption occurs at 308 nm, in good agreement with the experimental value of ca. 313 nm reported by Chevalier et al. \cite{chevalier2013esipt} in methylcyclohexane. 

At the Franck–Condon (FC) point of the ground state (\textit{cis}-)enol tautomer, the first two excited states, S$_1$ and S$_2$, are close in energy, with S$_1$ exhibiting bright $\pi\pi^*$ character and S$_2$ being dark with predominantly $n\pi^*$ character. Due to the near-degeneracy and the vibronic coupling of these states, even slight distortions of the molecular geometry can lead to mixing or even re-ordering of the states, as also observed by Nag et al. \cite{nag2021ultrafast}.
Owing to vibronic coupling between the first two excited states, the \say{dark} state can acquire intensity from the \say{bright} state when the molecular geometry is displaced along specific normal modes away from the FC region.
Due to this intensity borrowing both states are photo-excited and must be considered as initial states in the dynamics simulations.
In addition, the absorption spectrum was decomposed into the contribution from individual diabatic states (i.e., $\pi\pi^*$ and $n\pi^*$ character at the FC point) and reported in Figure S1. 


\subsection{ESIPT coordinate and competitive pathways characterisations}
\begin{figure}
    \centering
    \includegraphics[width=3.25in]{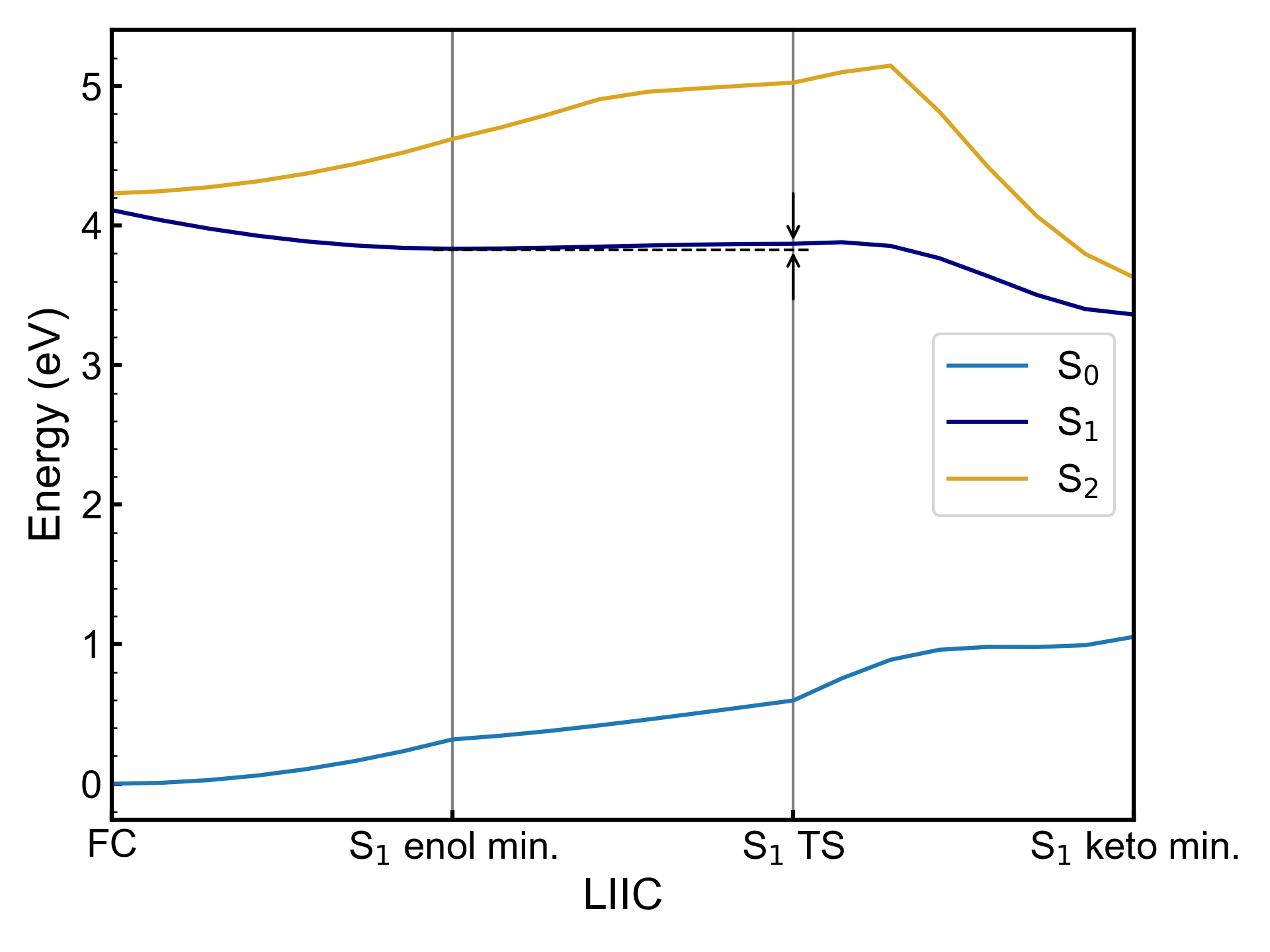}
    \caption{Linear interpolation in internal coordinates (LIIC) along the critical points that characterise the ESIPT coordinate at TD-PBE0/cc-pVDZ level of theory. The energy barrier on the S$_1$ potential energy surface between the ($cis$-)enol minimum to the TS is indicated by the two arrows.}
    \label{fig:liic_cc-pvdz}
\end{figure}

The ESIPT coordinate was characterised by locating key points on the potential energy surfaces and interpolating between them to obtain the energy profile along a collective coordinate that connects them. The structures considered include the FC point, the S$_1$ minima corresponding to the (\textit{cis}-)enol and keto tautomers (see Figure \ref{fig:3-HC_tautomers}), and the S$_1$ transition state (TS) connecting them. Linear interpolations in internal coordinates (LIIC) were performed between these critical points to map the ESIPT coordinate. The potential energy curves of the first three electronic states along these LIIC paths, computed at the TD-PBE0/cc-pVDZ level, are shown in Figure \ref{fig:liic_cc-pvdz}.
To assess the effect of basis set size, the critical points were also optimised with the cc-pVTZ and aug-cc-pVDZ basis set, and the LIIC profiles were recomputed. Expanding the basis from cc-pVDZ to cc-pVTZ or aug-cc-pVDZ has only a minor impact on the
potential energy curves along the ESIPT coordinate (see Figure S2). 


3-HC exhibits a small barrier of approximately 0.04 eV along the ESIPT coordinate on the S$_1$ potential energy surface, computed from the S$_1$ (\textit{cis}-)enol minimum to the TS on the same potential energy surface. In contrast, a significantly larger barrier of about 0.97 eV is observed along the same coordinate on the S$_2$ state potential energy surface, in qualitative agreement with the 0.95 eV barrier reported by Nag et al. \cite{nag2021ultrafast} at the TD-B3LYP/6-311++G** level. It is noted that optimising the ESIPT TS on the S$_2$ state gives a barrier of 0.46 eV, which is still approximately ten times higher than the one on the S$_1$ surface. In the ground-state, the keto form is not stable and all the optimisation attempts on this potential energy surface ended in the enol form. Consequently, the gradient at the FC point on the S$_1$ surface, together with the modest energy barrier along the proton-transfer coordinate, is expected to drive the system toward enol-to-keto tautomerization. In view of the substantially higher barrier on S$_2$, only a minor fraction of photoexcited molecules is anticipated to undergo ESIPT on this surface, as will be confirmed by the subsequent analysis of the non-adiabatic dynamics simulations (see below).

Previous studies \cite{chevalier2013esipt,perveaux2017fast} suggested that other relaxation processes may compete with proton transfer following photoexcitation of 3-HC, potentially explaining the double time constant observed for the ESIPT reaction. To investigate this, we first located a minimum-energy conical intersection (MECI) between the S$_1$ and S$_2$ states near the FC point, which was previously identified by Perveaux et al. \cite{perveaux2017fast} as a possible origin of the second ESIPT time constant.
This MECI adopts a planar geometry (C$_s$ symmetry, as the stationary points located and discussed so far) and is structurally very similar to the FC point, differing mainly by an increase of approximately 0.09 \r{A} in the $d(\mathrm{H}-\mathrm{O_a})$ distance. Notably, we find the MECI to lie 0.10 eV below the FC point, in agreement with Anand et al. \cite{anand2020excited}, suggesting that it is readily accessible to the nuclear wave packet following photoexcitation.
Perveaux et al. \cite{perveaux2017fast} reported that, possibly due to a different optimisation algorithm, the corresponding conical intersection lies slightly above the FC point, yet still confirms the accessibility of the S$_1$/S$_2$ seam in the vicinity of the FC point.

\begin{figure}
    \centering
    \includegraphics[width=3.25in]{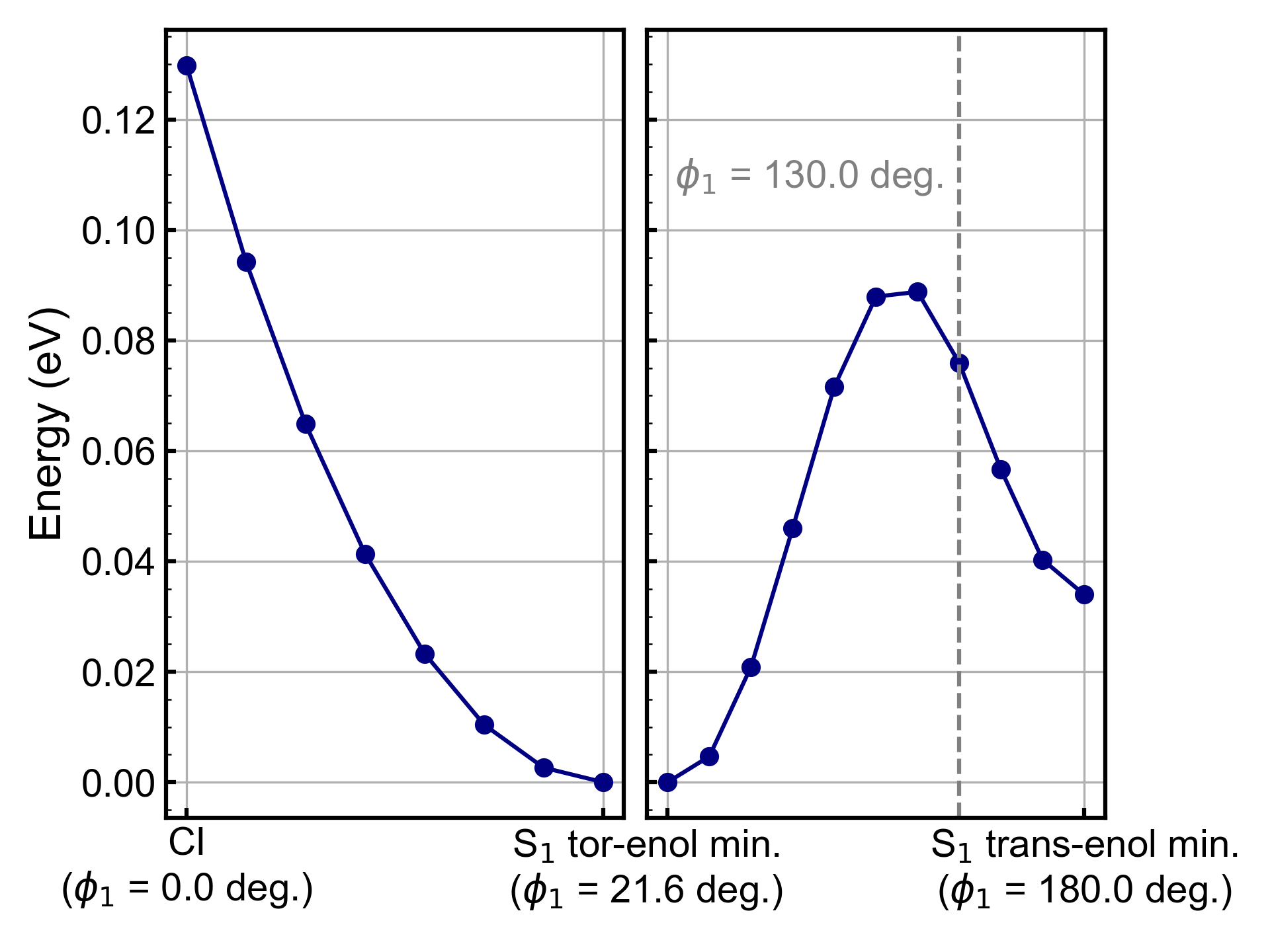}
    \caption{Linear interpolation in internal coordinates (LIIC) between the MECI and the S$_1$ torsional minimum (left) and relaxed scan from the S$_1$ torsional minimum to the planar \textit{trans}-enol conformer (right). Both the energy profiles were obtained at TD-PBE0/cc-pVDZ level.}
    \label{fig:liic_and_scan_torsion}
\end{figure}

After crossing the MECI seam, the S$_1$ state, now $n\pi^*$ in character, exhibits two symmetrically equivalent minima along the torsional coordinate defined by a $\phi_1$ dihedral angle at approximately $\pm 21.6$ degrees.
The loss of C$_s$ symmetry can be due to the pseudo-Jahn-Teller effect, arising from the vibronic coupling of the two states in the region of (near-)degeneracy \cite{bersuker2013pseudo}.
We refer to these stationary points as S$_1$ \say{torsional} minima (or in brevity, \textit{tor}-enol). To characterise the energy profile along this pathway, we performed a LIIC from the MECI to one of the two symmetrically equivalent torsional minima, followed by a relaxed scan from this minimum to the planar \textit{trans}-enol conformer (i.e., when $\phi_1 = \pm$180.0 degrees). The resulting S$_1$ energy curves are shown in Figure \ref{fig:liic_and_scan_torsion}. An exactly symmetric surface is expected when moving in the opposite (negative) direction along the $\phi_1$ torsion due to the symmetry of the system. From the MECI, the system can relax along $\phi_1$ to reach the torsional minimum without encountering any barrier. From this minimum, a modest barrier of 0.09 eV must be overcome to reach the \textit{trans}-enol conformer, in line with previous calculations \cite{perveaux2017fast}. Given the energy available upon vertical photoexcitation, the system is expected to explore this torsional coordinate, as will be confirmed by the non-adiabatic dynamics simulations.

\subsection{Non-adiabatic dynamics simulations}

\subsubsection{Electronic state populations}
\begin{figure}
    \centering
    \includegraphics[width=3.25in]{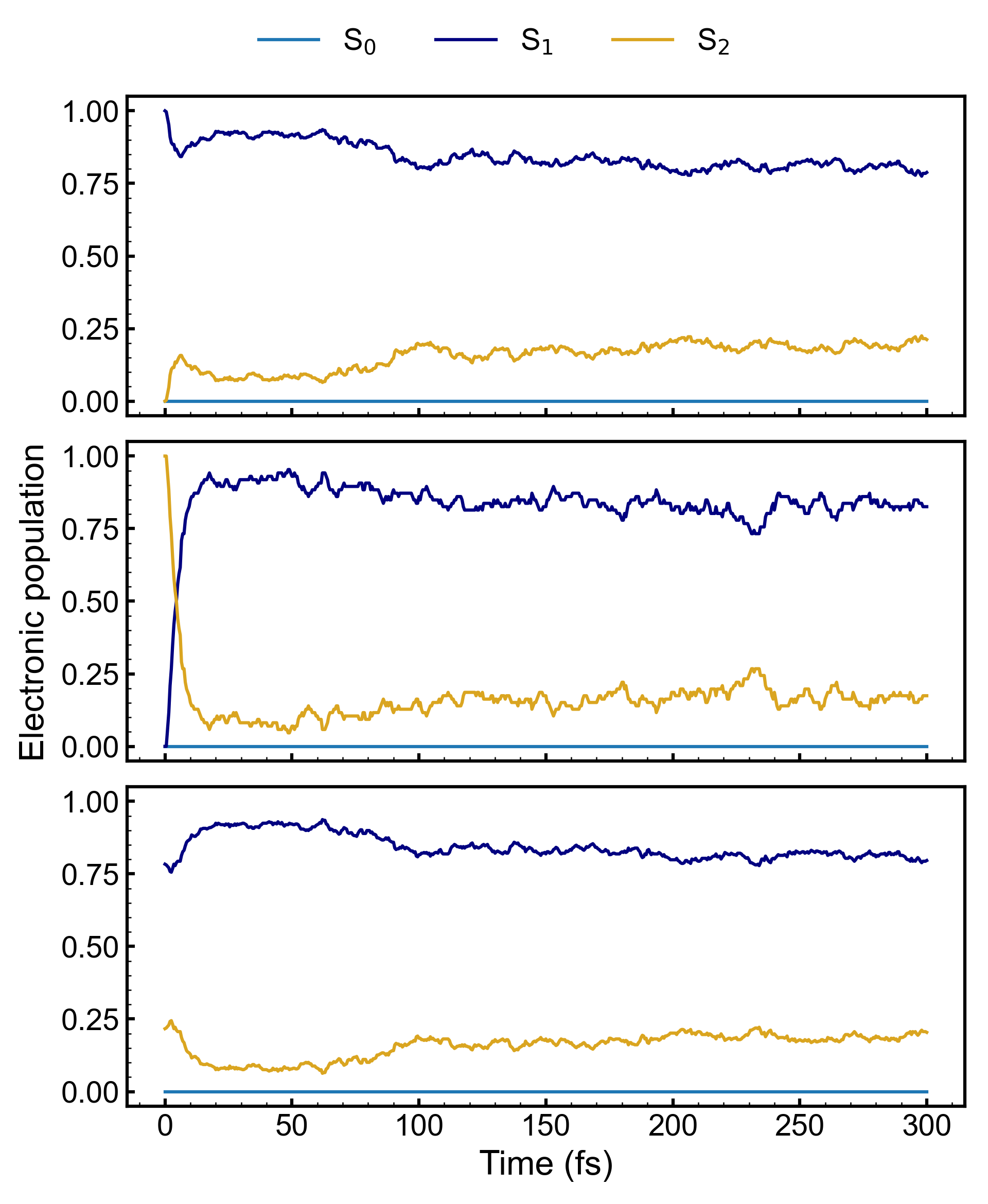}
    \caption{Electronic state population evolutions upon excitation to the S$_1$ (top) and S$_2$ (middle) electronic state. Combined S$_1$ and S$_2$ ensembles (bottom).}
    \label{fig:3-HC_sharc_elec_pop_10000initconds}
\end{figure}

We begin the discussion of the coupled electron–nuclear dynamics by analyzing the electronic state population evolutions, shown in Figure \ref{fig:3-HC_sharc_elec_pop_10000initconds} for each ensemble separately and for the combined trajectories. The convergence of the electronic populations over time was verified and is reported in Figure S3. 
The time evolution of the electronic populations for the full ensemble of trajectories is also reported in terms of the diabatic basis of $\pi\pi^*$ and $n\pi^*$ states at the ($cis$-)enol FC point in Figure S4. 
For the ensemble initialised in the S$_2$ state, the population of the initially active state decreases rapidly within the first 25 fs, driven by the narrow energy gap between S$_1$ and S$_2$ and the accessibility of the MECI near the FC region. A similar mechanism explains the initial depopulation of S$_1$ in favor of S$_2$ when the dynamics starts on the S$_1$ state, although this occurs to a lesser extent. 
The geometries at which S$_1$/S$_2$ hops occur during the first 25 fs predominantly adopt the (\textit{cis}-)enol conformation and are structurally similar to the MECI located near the FC point (see Figure S5 
), confirming that this intersection is readily accessible upon photoexcitation.

After the first 50 fs, the population transfer reaches an equilibrium in which the electronic populations are roughly constant in time. No population transfer from the excited states to the ground state is observed within the simulation time. It is consistent with the experimentally observed later radiative decay from the excited state to the ground state through fluorescence measurements\cite{klymchenko20043,giordano2023tuning}.

\subsubsection{Typical ESIPT trajectory and key degrees of freedom}
\begin{figure}
    \centering
    \includegraphics[width=3.25in]{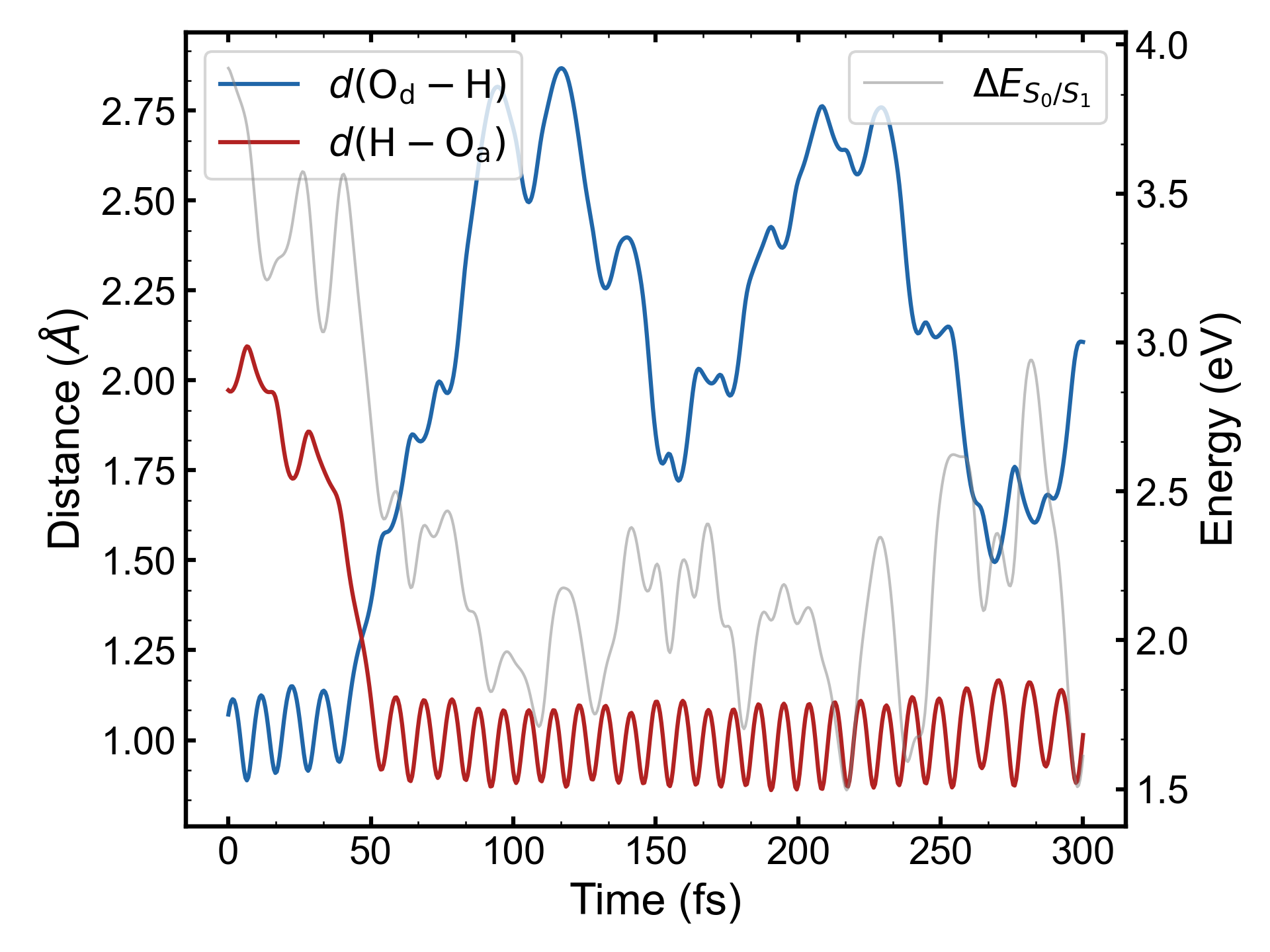}
    \caption{Key bond lengths and S$_1$/S$_0$ energy difference along a typical reactive trajectory observed in the simulations.}
    \label{fig:typical_reactive_trj}
\end{figure}

The key bond lengths and S$_1$/S$_0$ energy difference along a typical reactive trajectory are depicted in Figure \ref{fig:typical_reactive_trj}.
At early times, the molecule evolves within the reactant well, as evidenced by the rapid fluctuations of the donor oxygen-hydrogen distance (blue curve) at a frequency characteristic of the oxygen–hydrogen stretching mode (ca. 3579 cm$^{-1}$, corresponding to a period of ca. 9.3 fs). In this trajectory, proton transfer, defined by the proton-acceptor oxygen distance (red curve) becoming shorter than the donor oxygen-proton distance, occurs around 50 fs after photo-excitation. This transfer is accompanied by a marked decrease in the ground/excited state energy gap (grey curve), reflecting the large Stokes shift typical of ESIPT chromophores. After the proton transfer, the system resides in the product tautomer well, where the hydrogen–acceptor oxygen distance exhibits the characteristic oxygen-hydrogen bond stretching frequency.
The low frequency oscillation observed in the donor oxygen-hydrogen distance likely arises from the in-plane bending motions, involving the carbonyl and hydroxyl functional groups, which periodically drive the hydrogen atom back toward the donor oxygen. The normal modes (calculated at the keto tautomer S$_1$ equilibrium geometry) associated with this motion have frequencies of 263 and 316 cm$^{-1}$, corresponding to periods of approximately 106 to 126 fs, in qualitative agreement with the oscillation period observed following proton transfer.

\begin{figure}
    \centering
    \includegraphics[width=3.25in]{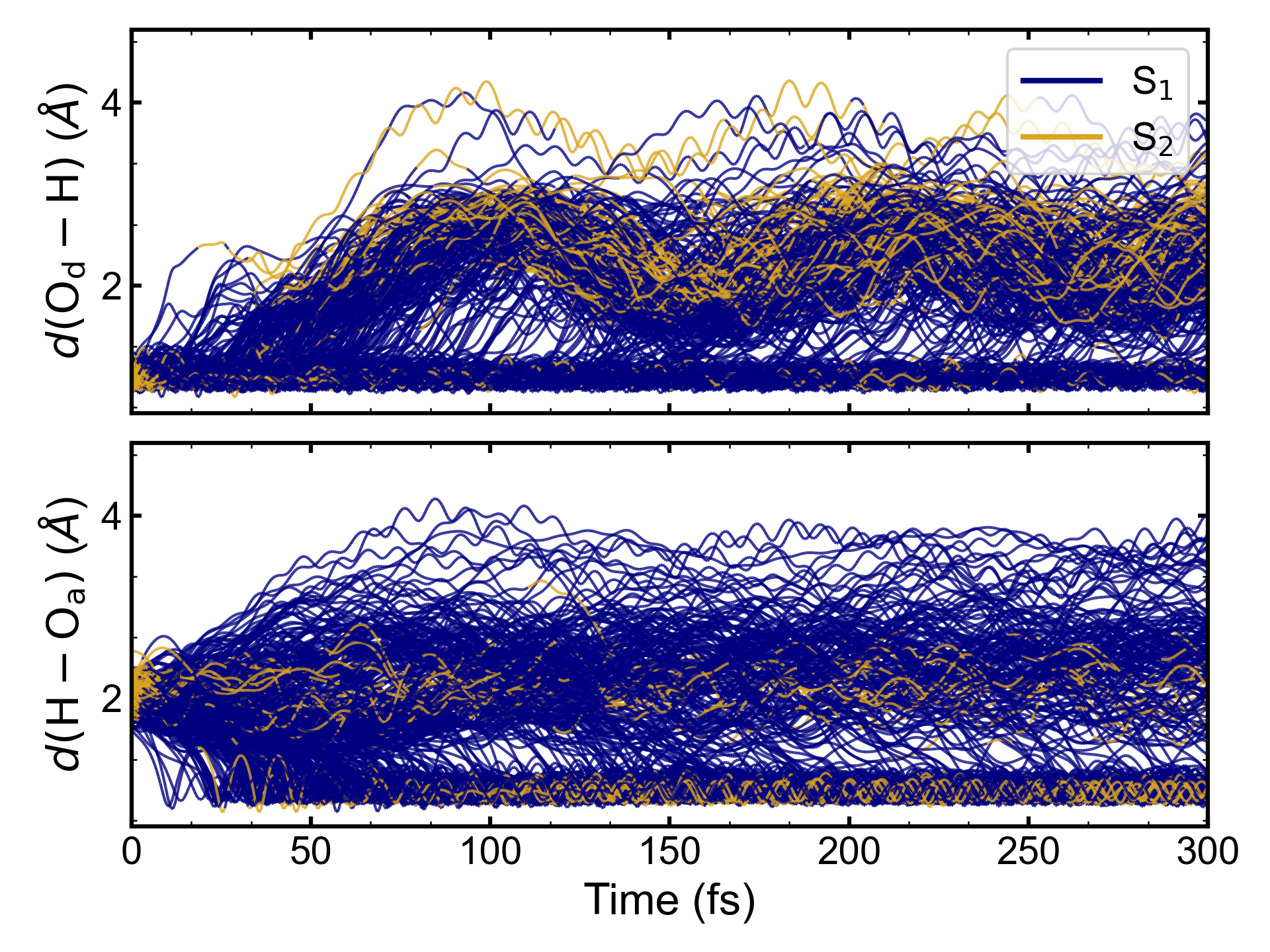}
    \caption{The key oxygen-hydrogen distances involved in the description of the ESIPT process in time along all trajectories. The color code indicate the active state.}
    \label{fig:3-HC_sharc_hbonds_10000initconds}
\end{figure}

Next, to gain statistical insight into the ESIPT process and possible competing pathways, the key internal coordinates are examined for the ensemble of mixed quantum–classical surface-hopping trajectories. Figure \ref{fig:3-HC_sharc_hbonds_10000initconds} shows the evolution along all trajectories of the donor oxygen–hydrogen bond distance, $d(\mathrm{O_d}-\mathrm{H})$, and the hydrogen–acceptor oxygen bond distance, $d(\mathrm{H}-\mathrm{O_a})$, which directly describe the proton transfer reaction. The $d(\mathrm{O_d}-\mathrm{H})$ distances are initially distributed around the equilibrium value of 0.98 \r{A} at the FC point, but after only 25 fs part of the trajectories exhibits the elongation of the bond distance in favour of the formation of the proton-acceptor oxygen bond, testified by the swarm of trajectories showing the shortening of the $d(\mathrm{H}-\mathrm{O_a})$ distance from the average value of 2.00 \r{A} to 0.98 \r{A}. In addition to the shortening of $d(\mathrm{H}-\mathrm{O_a})$, we also observe an increase after 50 fs for another portion of the analysed trajectories caused by the torsion of the hydroxyl group around the dihedral angle $\phi_1$.

In nearly all cases where $d(\mathrm{O_d}-\mathrm{H})$ increases while $d(\mathrm{H}-\mathrm{O_a})$ decreases, leading to the proton remaining bound to the acceptor oxygen, the system resides in the S$_1$ electronic state. This is consistent with the lower barrier height along the ESIPT coordinate on S$_1$ (about 0.42 eV 
lower than on S$_2$). Consequently, upon excitation to S$_2$, the system relaxes efficiently to S$_1$ before proton transfer takes place, in agreement with Nag et al. \cite{nag2021ultrafast}.
This finding rules out ESIPT occurring directly on the S$_2$ surface as the origin of the second (slower) ESIPT time constant, at least within the time window accessible to our simulations.

\subsubsection{Kinetic modelling}
\begin{figure}
    \centering
    \includegraphics[width=3.25in]{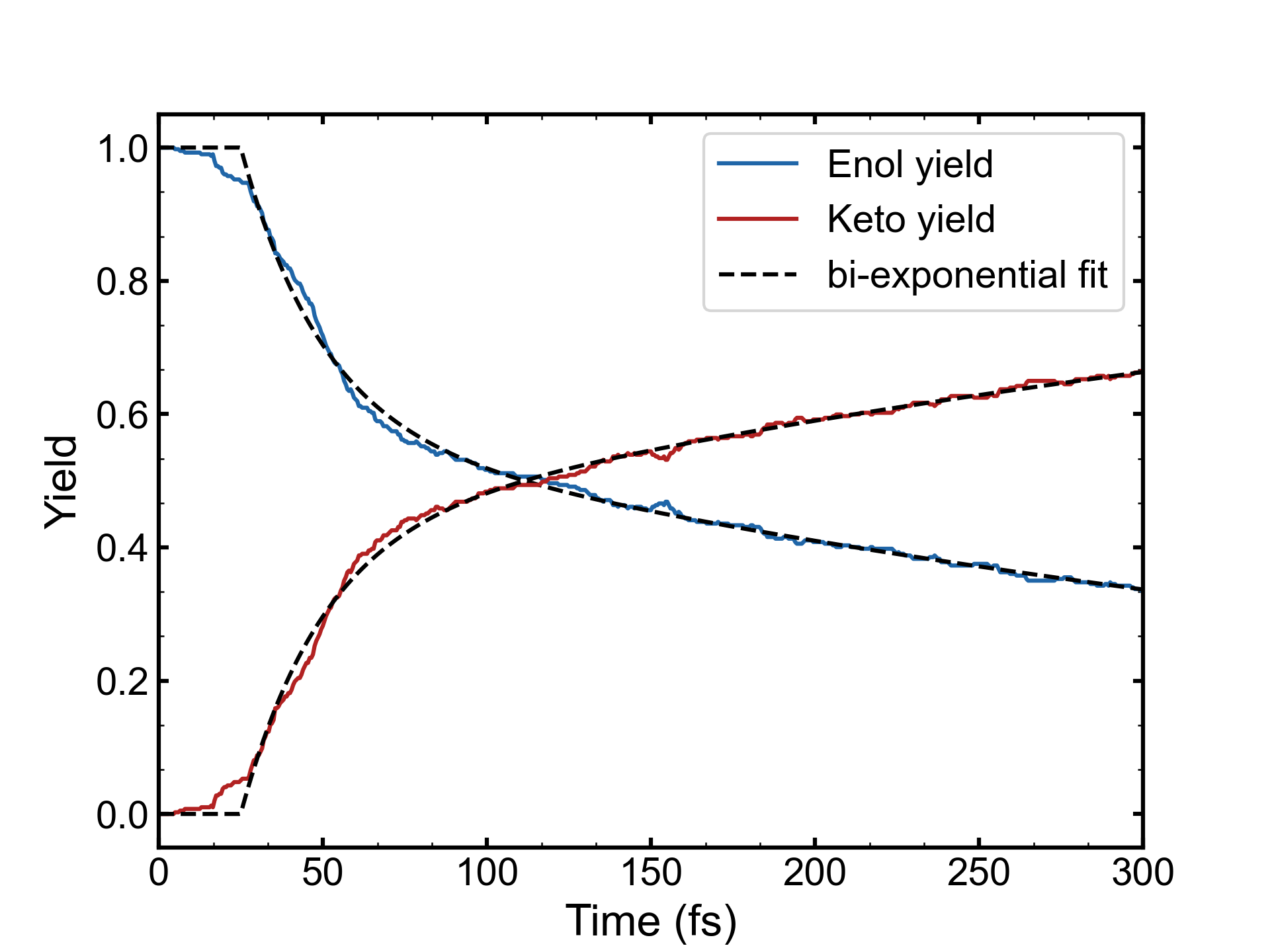}
    \caption{Time-dependent yields of the reactant (enol, in blue) and product (keto, in red) forms. The black dashed line represents the bi-exponential fit of the reactant and product populations in time.}
    \label{fig:yields_from_both_ens}
\end{figure}

The yield of the enol and keto forms was monitored in time using the key distances described above. At each time step, the molecular structure visited along the trajectory is classified as the enol tautomer if $d(\mathrm{O_d}-\mathrm{H}) \le d(\mathrm{H}-\mathrm{O_a})$ and as the keto tautomer if $d(\mathrm{O_d}-\mathrm{H}) > d(\mathrm{H}-\mathrm{O_a})$ distance. Figure \ref{fig:yields_from_both_ens} shows the population dynamics of both tautomers obtained from the combined analysis of the two ensembles of trajectories.
Proton migration was observed in 69.8\% of the trajectories.
The convergence of the tautomeric yields over time for each ensemble individually is reported in Figure S6. 

The time evolution of the enol population was fitted using a bi-exponential function with latency of the form

\begin{equation}
    Y(t) = 
    \begin{cases}
        1, \quad t < t_0 \\
        \delta_1 \exp \left( -\frac{(t - t_0)}{\tau_1} \right) + \delta_2 \exp \left( -\frac{(t - t_0)}{\tau_2} \right) + C, \quad t \ge t_0
    \end{cases}
\end{equation}
where $\delta_1$ and $\delta_2$ are the fractions of population decaying with time constants $\tau_1$ and $\tau_2$, respectively, $t_0$ is the latency time, $C$ is the asymptotic population. The optimised parameters of the bi-exponential fit are reported in Table \ref{tab:fit_biexp}. A reliable description of the reactant decay requires a bi-exponential function rather than a mono-exponential decay. A comparison between mono- and bi-exponential fits is shown in Figure S7 
and Table S1. 

\begin{table}[h]
    \centering
    \begin{tabular}{ccc}
    \hline
    \multicolumn{3}{c}{Fit parameters} \\
    \hline
    $t_0$ (fs) & $\delta_1$ & $\delta_2$ \\
    \hline
    24.97 (0.20) & 0.42 (0.01) & 0.58 (0.14) \\
    \hline
    $\tau_1$ (fs) & $\tau_2$ (fs) & $C$ \\
    \hline
    24.82 (0.84) & 510.92 (162.79) & 0.00 (0.13) \\
    \hline
    \end{tabular}
    \caption{Optimised parameters from the fit of the enol yield time evolution with a bi-exponential decay function. In parentheses one standard deviation errors on the fit parameters are reported.}
    \label{tab:fit_biexp}
\end{table}

From the bi-exponential fit, two time constants for the ESIPT process were extracted: a fast component on the sub-100 fs timescale (24.82 fs), consistent with previous non-adiabatic dynamics studies \cite{nag2021ultrafast}, and a slower component of approximately 0.51 ps. Experimentally, a second ESIPT time-constant of ca. 5.5 ps was previously found for the same system in methylcyclohexane \cite{chevalier2013esipt}.
To obtain reliable estimate of the uncertainty on the predicted slower component, which is beyond the simulated time scale, a bootstrap procedure was employed. From the set of all the trajectories, 200 set made by 100 trajectories were sampled with replacement and from each set we obtained the $\tau_2$ from the fit of the enol and keto yields. The 3rd and 97th percentiles are 0.11 and 1.15 ps, respectively. Despite the large noise, our identification of an additional picosecond phenomenon is in qualitative agreement with previous experiments \cite{chevalier2013esipt}.
The results from the non-adiabatic dynamics simulations in the gas phase are compared with measurements in methylcyclohexane. Due to the aprotic non-polar nature of the solvent, only a limited effect on the time scales of the investigated processes is expected. This is confirmed by a comparison between the gas-phase LIIC and the one in methylcyclohexane (Figure S8). 
However, subtle solvation effects could explain a quantitative discrepancy on the picosecond time scale.

\subsubsection{Origin of the second ESIPT time constant and competing pathway}
The enol and keto yields in time from the ensembles of trajectories that started from the excited-states S$_1$ and S$_2$ were analysed separately (Figure S9 and S10 in the SI). 
For both ensembles, two ESIPT time scales are observed and they are indistinguishable within the error (see Table S2 and Table S3), 
indicating that the two ESIPT time scales do not originate from the different initial electronic state upon excitation.

\begin{figure}
    \centering
    \includegraphics[width=3.25in]{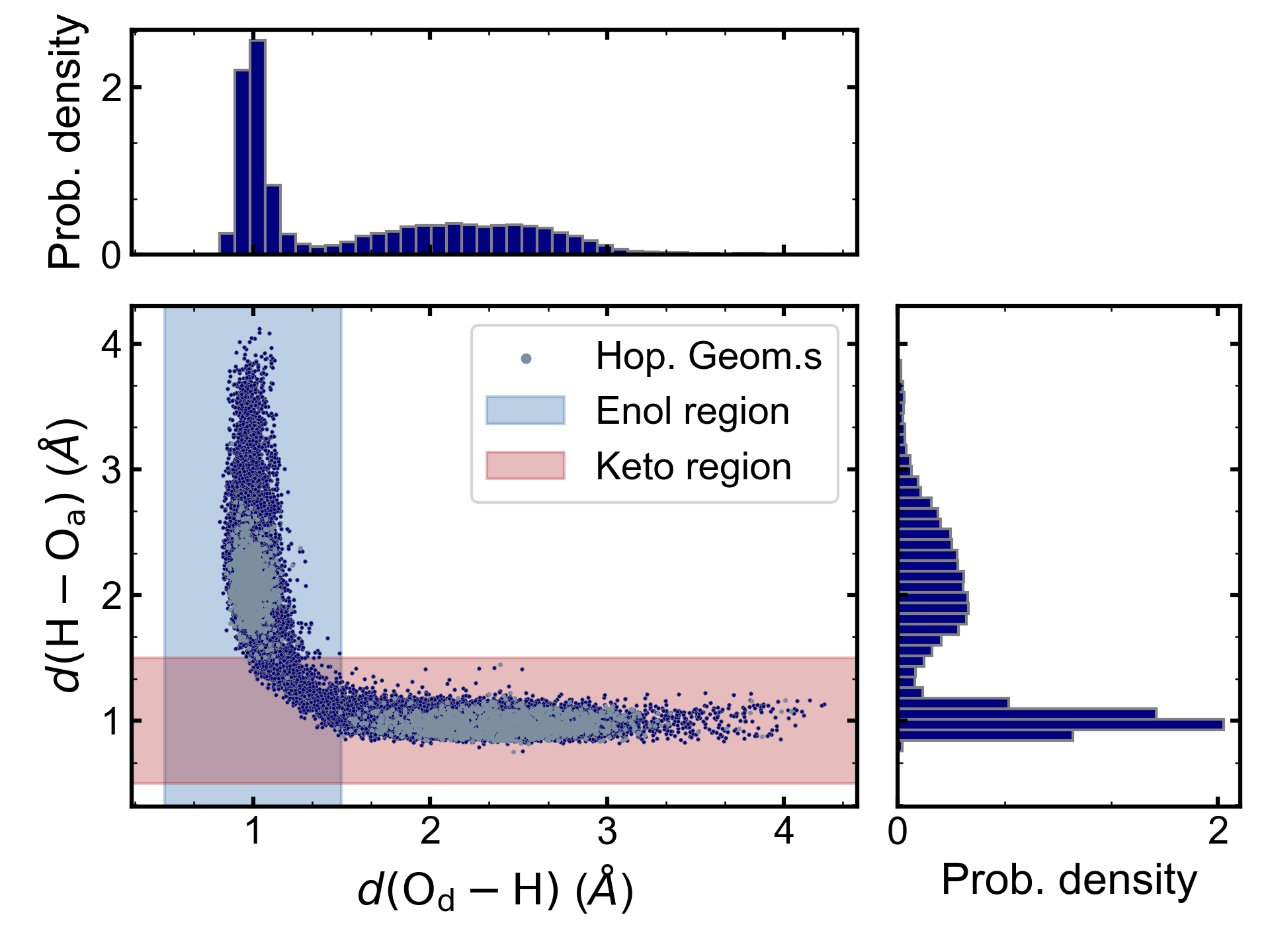}
    \caption{Scatter plot of the $d(\mathrm{O_d}-\mathrm{H})$ and $d(\mathrm{H}-\mathrm{O_a})$ distances visited by the 3-HC along the surface hopping trajectories.}
    \label{fig:3-HC_sharc_distances_10000initconds}
\end{figure}

\begin{figure}
    \centering
    \includegraphics[width=3.25in]{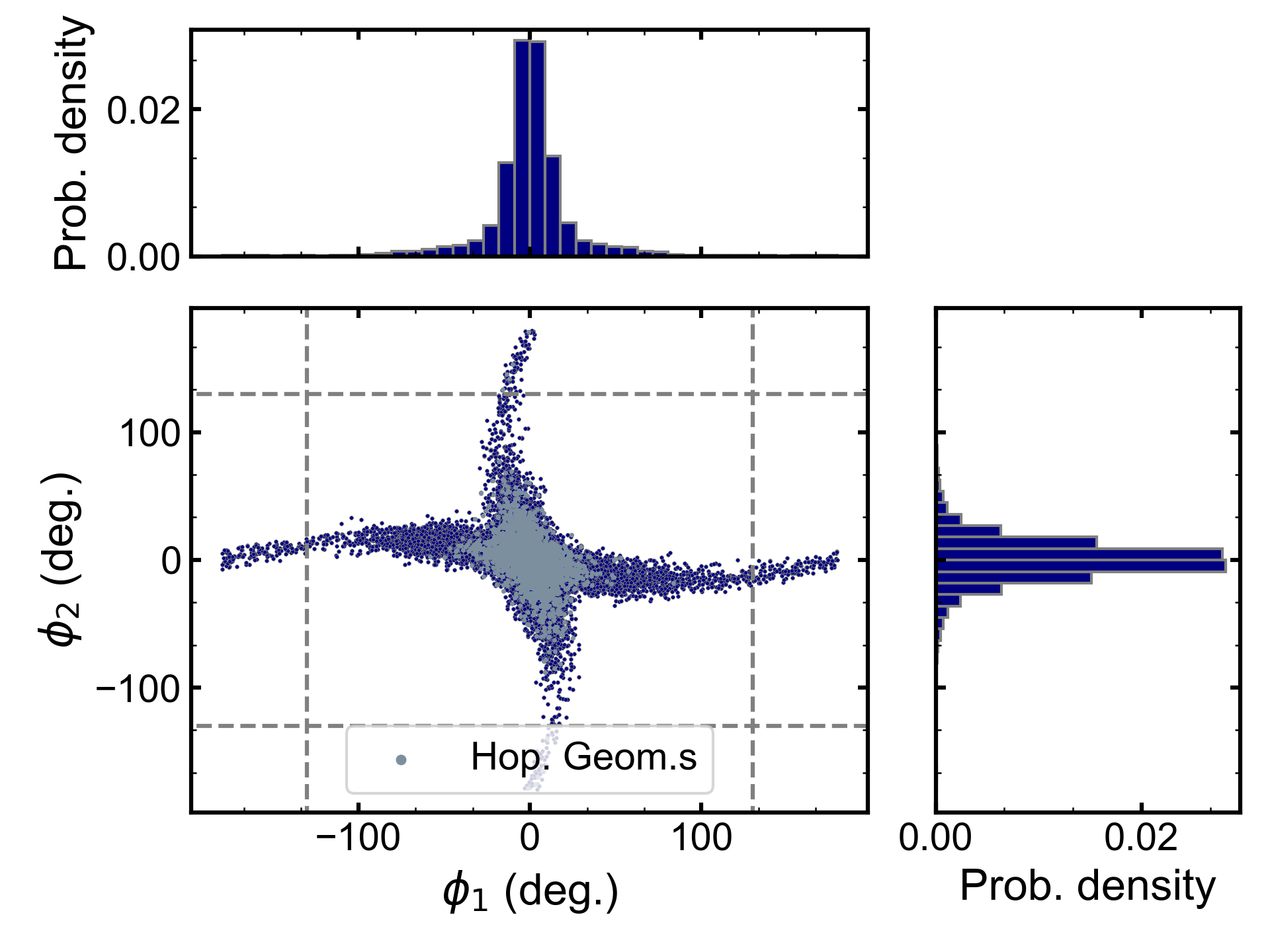}
    \caption{Scatter plot of the $\phi_1$ and $\phi_2$ dihedral angles visited by the 3-HC along the surface hopping trajectories.}
    \label{fig:3-HC_sharc_dihedrals_10000initconds}
\end{figure}

To understand the origin of the two time constants, we inspected the distributions of the donor oxygen-hydrogen and hydrogen-acceptor oxygen distances, and of the dihedral angles $\phi_1$ and $\phi_2$ (in Figures \ref{fig:3-HC_sharc_distances_10000initconds} and \ref{fig:3-HC_sharc_dihedrals_10000initconds}, respectively). The hopping geometries are highlighted with a different colour in both the internal coordinate subspaces. 
The tails of the distance distributions (i.e., for $d(\mathrm{O_d}-\mathrm{H})$, $d(\mathrm{H}-\mathrm{O_a}) > 3.4$ \r{A}) in Figure \ref{fig:3-HC_sharc_distances_10000initconds} suggest the formation of the \textit{trans} isomer of both enol and keto forms. It is also possible to clearly distinguish the hopping geometries into enol-type and keto-type structures from the distance subspace.
From Figure \ref{fig:3-HC_sharc_dihedrals_10000initconds} it is possible to appreciate and confirm the exploration of the configurational space associated with the \textit{trans} form of the enol (i.e., when $|\phi_1| > 130.0$ degrees) and of the keto (i.e., when $|\phi_2| > 130.0$ degrees). The $130.0$ degrees threshold value was selected based on the results of the scan along the hydrogen torsion coordinate (see Figure \ref{fig:liic_and_scan_torsion}). At $\phi_1 \simeq 130.0$ degrees the system fully overcame the energy barrier to reach the \textit{trans}-enol conformer.

Statistically, 54.4\% of trajectories originating from the (\textit{cis}-)enol conformer reach the region of the unreactive S$_1$ torsional minimum (with $n\pi^*$ character and $|\phi_1|$ around 21.6 degrees).
Among these, 10.2\% proceed further to undergo the full (\textit{cis}-)enol to \textit{trans}-enol transition ($|\phi_1| > 130.0$ degrees), i.e., crossing the barrier connecting the torsional minimum to the \textit{trans}-enol conformation.
These observations indicate that, after crossing the S$_1$/S$_2$ seam near the FC point, the system extensively explores the configurational space associated with out-of-plane hydrogen torsion. In some cases, this leads to formation of the \textit{trans}-enol conformer, representing a competitive pathway to ESIPT.
Notably, the amplitude associated with the slower component of the bi-exponential fit is ca. 58.0\% which is comparable to the fraction of the trajectories (54.4\%) that visit the region of the S$_1$ torsional minimum, supporting the assignment of the longer time constant to this torsional dynamics.

\begin{figure}
    \centering
    \includegraphics[width=3.25in]{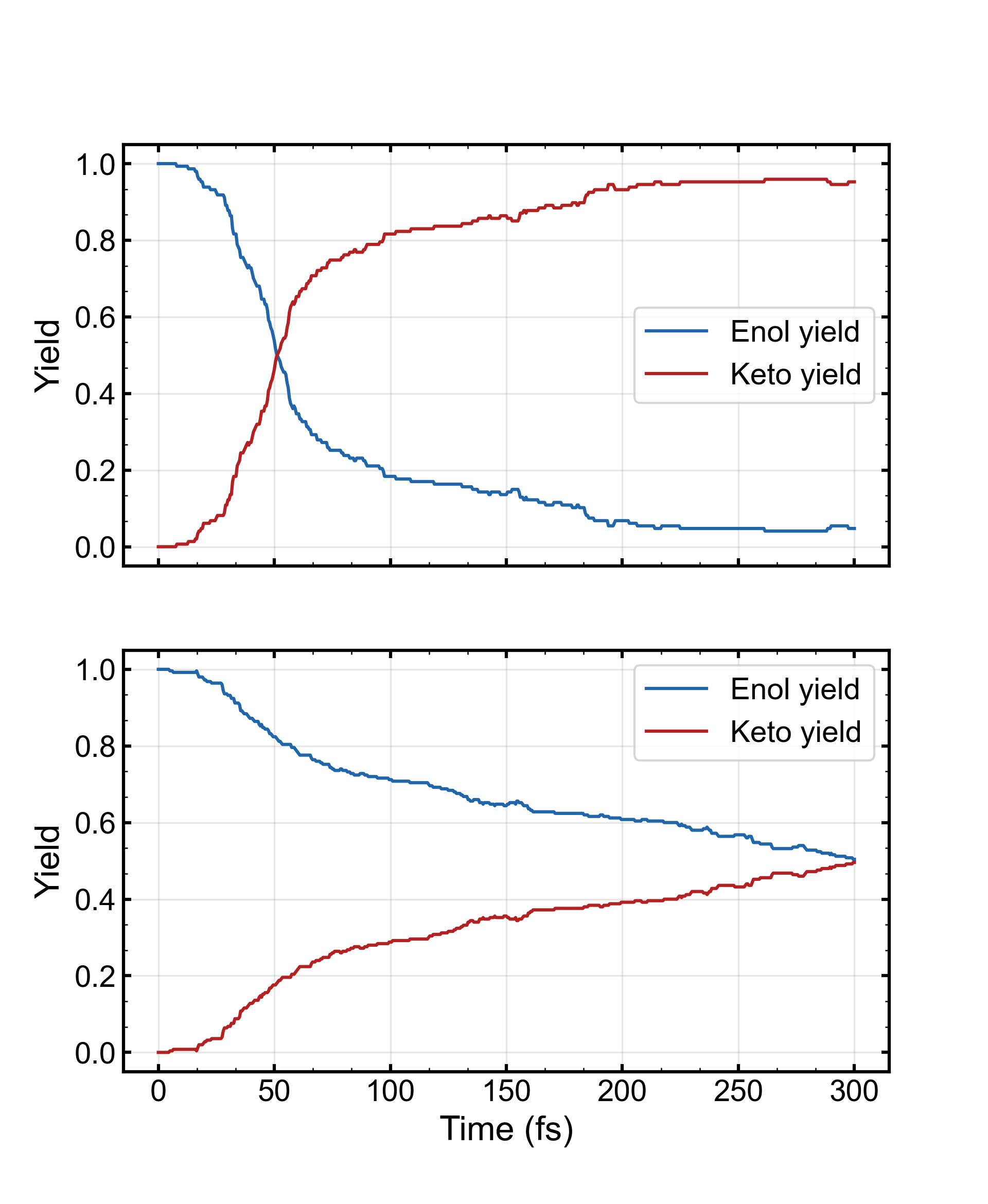}
    \caption{Time-dependent yield of the reactant (enol, in blue) and product (keto, in red) forms obtained from (top) the subset of trajectories that do not display a torsion along the $|\phi_1|$ coordinate more than 21.6 degrees and (bottom) the subset of trajectories that meet at least in one frame the condition $|\phi_1| \ge 21.6$ degrees.}
    \label{fig:less_greater_216}
\end{figure}

To test this hypothesis, we analyzed the time-dependent yields of the ESIPT reactant (enol) and product (keto) for two subsets of trajectories. The first subset comprises trajectories in which the $|\phi_1|$ torsion never exceeds 21.6 degrees, while the second subset includes trajectories in which $|\phi_1| > 21.6$ degrees in at least one frame.
As expected, the subset without significant torsion along $\phi_1$ exhibits rapid enol-to-keto conversion, whereas the subset visiting the S$_1$ torsional minimum shows a delayed proton transfer (Figure \ref{fig:less_greater_216}). These results support the idea that out-of-plane hydrogen torsion, readily accessible after crossing the S$_1$/S$_2$ seam near the FC region, modulates the ESIPT timescales and gives rise to the second slower proton transfer time constant.
To assess whether the observed differences between the two trajectory subsets arise from structural features inherent to the initial positions or momenta, we performed a linear discriminant analysis (LDA) \cite{fisher1936use,hastie2009elements} on their respective initial conditions. The LDA results indicate a partial separation of the initial conditions along a discriminant axis dominated by the hydroxyl group’s out-of-plane motion (Figure S11). 

\subsubsection{Global reaction network}

The analyses linked the slower ESIPT time constant to out-of-plane hydrogen torsion and visits to the S$_1$ torsional minimum. The system’s excited-state dynamics involve a broad network of conformational changes and tautomer interconversions. To check the dominant collective motions driving the excited-state evolution, we performed a principal component analysis (PCA) \cite{pearson1901liii} on the full ensemble of trajectories (see SI, Section S11). 
It confirms that the previously identified critical structures represent the main basins/critical points of the excited-state dynamics.
Accordingly, all structures visited along the trajectories were clustered into four categories based on their geometry:

\begin{itemize}
    \item \textit{cis}-enol ($d(\mathrm{O_d}-\mathrm{H}) \le d(\mathrm{H}-\mathrm{O_a})$ and $|\phi_1| <$ 21.6 degrees)
    \item \textit{tor}-enol ($d(\mathrm{O_d}-\mathrm{H}) \leq d(\mathrm{H}-\mathrm{O_a})$ and 21.6 degrees $\le |\phi_1| <$ 130.0 degrees)
    \item \textit{trans}-enol ($d(\mathrm{O_d}-\mathrm{H}) \leq d(\mathrm{H}-\mathrm{O_a})$ and $|\phi_1| \ge$ 130.0 degrees)
    \item keto ($d(\mathrm{O_d}-\mathrm{H}) > d(\mathrm{H}-\mathrm{O_a})$)
\end{itemize}

\begin{figure}
    \centering
    \includegraphics[width=3.25in]{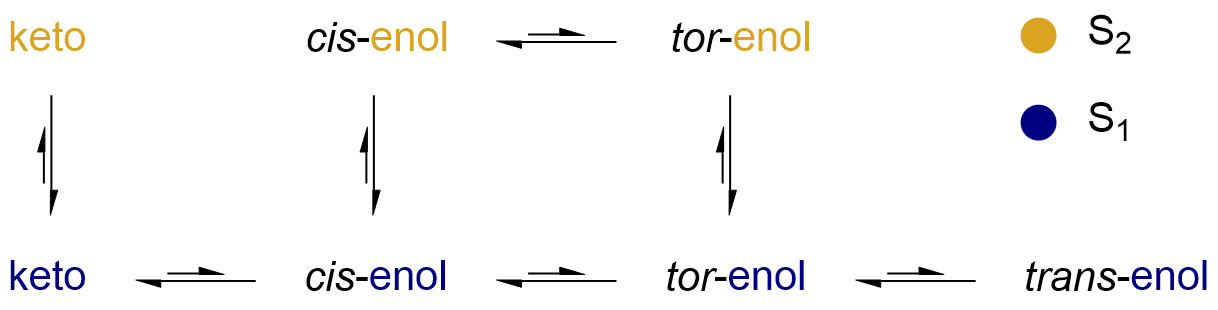}
    \caption{Reaction network.}
    \label{fig:rea_network}
\end{figure}

Since each conformer/tautomer can be populated in either the S$_1$ or S$_2$ electronic state, this leads to a total of eight distinct states. Every frame of the ensemble of trajectories was assigned to one of these states, and the transitions between them were counted. The resulting transition matrix (with normalised rows) is reported in Figure S14. 
From this state-to-state connectivity, we constructed the complete reaction network shown in Figure \ref{fig:rea_network}.

It is possible to observe a population transfer between the electronic states S$_1$ and S$_2$ when the system reaches the proton transfer product, i.e., the keto species. Indeed, we found several hopping structures and two non-planar MECI that share structural features with the keto tautomer (see Figure S15). 
The first keto-type MECI is located ca. 0.18 eV above the keto minimum on the S$_1$ surface and it results accessible to the system after the proton is transferred. The second keto-type MECI is characterised by the torsion of the hydrogen that was previously transferred and it is located ca. 0.52 eV above the keto minimum on the S$_1$ surface. Even if less accessible with respect to the previous one, this observation is consistent with the finding of several crossing geometries from the mixed quantum classical simulations adopting a \textit{trans}-keto conformation.

Finally, this extended picture of the dynamics of the 3-HC system upon excitation to S$_1$ or S$_2$ corroborates the presence of a competitive pathway to ESIPT, associated with the out-of-plane motion of the hydrogen atom rather than its direct transfer between the two oxygen atoms.

\section{Conclusions}
\label{sec:conclusions}
The present work investigated the photodynamics of the 3-HC molecule, with a focus on the excited-state intramolecular proton transfer and the possible competitive relaxation pathways. The coupled electron–nuclear dynamics of 3-HC following excitation to the S$_1$ and S$_2$ states were simulated using surface hopping with TD-PBE0 for the electronic structure.
Consistent with previous theoretical studies, our simulations indicate that ESIPT takes place on a sub-100 fs timescale. In addition, we directly observed a slower component of the proton transfer dynamics, giving rise to a second time constant, which was not reported in earlier simulations. The second ESIPT timescale predicted by our simulations is in qualitative agreement with experimental observations from time-resolved UV–vis spectroscopy in organic aprotic apolar solvents.

Our analysis of the critical points on the potential energy surfaces, together with the ensemble of mixed quantum-classical trajectories, reveals that the slower ESIPT component originates from competition with the out-of-plane torsional motion of the transferring proton, accessible after crossing the S$_1$/S$_2$ seam near the Franck-Condon region.
Ultrafast transient spectroscopic techniques may detect the competitive torsional out-of-plane motion corroborating the findings of our non-adiabatic dynamics simulations.
Finally, by clustering the visited geometries into representative conformational/tautomeric states and monitoring their interconversion, we constructed an explicit reaction network for 3-HC in the excited state. This network highlights the coexistence of the canonical ESIPT pathway with torsion-mediated channels, offering a unified mechanistic picture that reconciles the ultrafast and slower components of the ESIPT dynamics.

\section*{Conflicts of interest}
There are no conflicts to declare.

\section*{Acknowledgements}
The authors acknowledge Benjamin Lasorne for reading the original draft of the manuscript and for the stimulating discussion.
The simulations in this work were performed using HPC resources from CCIPL (Le center de calcul intensif des Pays de la Loire) and from GENCI-IDRIS (grant 2021-101353). The project is funded by the European Union (ERC, 101040356, ATTOP). Views and opinions expressed are those of the authors only and do not necessarily reflect those of the European Union or the European Research Council Executive Agency. Neither the European Union nor the granting authority can be held responsible for them.

\bibliographystyle{unsrt}  

\bibliography{references}

@article{weller1955fluoreszenz,
  title={{\"U}ber die Fluoreszenz der Salizyls{\"a}ure und verwandter Verbindungen},
  author={Weller, Albert},
  journal={Naturwissenschaften},
  volume={42},
  number={7},
  pages={175--176},
  year={1955},
  publisher={Springer}
}

@article{weller1956innermolekularer,
  title={Innermolekularer protonen{\"u}bergang im angeregten zustand},
  author={Weller, Albert},
  journal={Zeitschrift f{\"u}r Elektrochemie, Berichte der Bunsengesellschaft f{\"u}r physikalische Chemie},
  volume={60},
  number={9-10},
  pages={1144--1147},
  year={1956},
  publisher={Wiley Online Library}
}

@article{otterstedt1973photostability,
  title={Photostability and molecular structure},
  author={Otterstedt, Jan-Erik A},
  journal={The Journal of Chemical Physics},
  volume={58},
  number={12},
  pages={5716--5725},
  year={1973},
  publisher={American Institute of Physics}
}

@article{sengupta1979excited,
  title={Excited state proton-transfer spectroscopy of 3-hydroxyflavone and quercetin},
  author={Sengupta, Pradeep K and Kasha, Michael},
  journal={Chemical Physics Letters},
  volume={68},
  number={2-3},
  pages={382--385},
  year={1979},
  publisher={Elsevier}
}

@article{sytina2008conformational,
  title={Conformational changes in an ultrafast light-driven enzyme determine catalytic activity},
  author={Sytina, Olga A and Heyes, Derren J and Hunter, C Neil and Alexandre, Maxime T and Van Stokkum, Ivo HM and Van Grondelle, Rienk and Groot, Marie Louise},
  journal={Nature},
  volume={456},
  number={7224},
  pages={1001--1004},
  year={2008},
  publisher={Nature Publishing Group UK London}
}

@article{tonge2009excited,
  title={Excited state dynamics in the green fluorescent protein},
  author={Tonge, Peter J and Meech, Stephen R},
  journal={Journal of Photochemistry and Photobiology A: Chemistry},
  volume={205},
  number={1},
  pages={1--11},
  year={2009},
  publisher={Elsevier}
}

@article{weinberg2012proton,
  title={Proton-coupled electron transfer},
  author={Weinberg, David R and Gagliardi, Christopher J and Hull, Jonathan F and Murphy, Christine Fecenko and Kent, Caleb A and Westlake, Brittany C and Paul, Amit and Ess, Daniel H and McCafferty, Dewey Granville and Meyer, Thomas J},
  journal={Chemical Reviews},
  volume={112},
  number={7},
  pages={4016--4093},
  year={2012},
  publisher={ACS Publications}
}

@article{jacquemin2014assessing,
  title={Assessing the importance of proton transfer reactions in DNA},
  author={Jacquemin, Denis and Zuniga, Jose and Requena, Alberto and C{\'e}ron-Carrasco, Jos{\'e} Pedro},
  journal={Accounts of Chemical Research},
  volume={47},
  number={8},
  pages={2467--2474},
  year={2014},
  publisher={ACS Publications}
}

@article{joshi2021excited,
  title={Excited-state intramolecular proton transfer: A short introductory review},
  author={Joshi, Hem C and Antonov, Liudmil},
  journal={Molecules},
  volume={26},
  number={5},
  pages={1475},
  year={2021},
  publisher={MDPI}
}

@article{zhao2012excited,
  title={Excited state intramolecular proton transfer (ESIPT): from principal photophysics to the development of new chromophores and applications in fluorescent molecular probes and luminescent materials},
  author={Zhao, Jianzhang and Ji, Shaomin and Chen, Yinghui and Guo, Huimin and Yang, Pei},
  journal={Physical Chemistry Chemical Physics},
  volume={14},
  number={25},
  pages={8803--8817},
  year={2012},
  publisher={Royal Society of Chemistry}
}

@article{sedgwick2018excited,
  title={Excited-state intramolecular proton-transfer (ESIPT) based fluorescence sensors and imaging agents},
  author={Sedgwick, Adam C and Wu, Luling and Han, Hai-Hao and Bull, Steven D and He, Xiao-Peng and James, Tony D and Sessler, Jonathan L and Tang, Ben Zhong and Tian, He and Yoon, Juyoung},
  journal={Chemical Society Reviews},
  volume={47},
  number={23},
  pages={8842--8880},
  year={2018},
  publisher={Royal Society of Chemistry}
}

@article{li2021progress,
  title={Progress in tuning emission of the excited-state intramolecular proton transfer (ESIPT)-based fluorescent probes},
  author={Li, Yonghao and Dahal, Dipendra and Abeywickrama, Chathura S and Pang, Yi},
  journal={ACS Omega},
  volume={6},
  number={10},
  pages={6547--6553},
  year={2021},
  publisher={ACS Publications}
}

@article{paterson2005mechanism,
  title={Mechanism of an exceptional class of photostabilizers: a seam of conical intersection parallel to excited state intramolecular proton transfer (ESIPT) in o-hydroxyphenyl-(1, 3, 5)-triazine},
  author={Paterson, Martin J and Robb, Michael A and Blancafort, Llu{\'\i}s and DeBellis, Anthony D},
  journal={The Journal of Physical Chemistry A},
  volume={109},
  number={33},
  pages={7527--7537},
  year={2005},
  publisher={ACS Publications}
}

@article{gong2016new,
  title={New ESIPT-inspired photostabilizers of two-photon absorption coumarin--benzotriazole dyads: From experiments to molecular modeling},
  author={Gong, Yulong and Wang, Zhenqiang and Zhang, Shengtao and Luo, Ziping and Gao, Fang and Li, Hongru},
  journal={Industrial \& Engineering Chemistry Research},
  volume={55},
  number={18},
  pages={5223--5230},
  year={2016},
  publisher={ACS Publications}
}

@article{kwon2011advanced,
  title={Advanced organic optoelectronic materials: Harnessing excited-state intramolecular proton transfer (ESIPT) process},
  author={Kwon, Ji Eon and Park, Soo Young},
  journal={Advanced Materials},
  volume={23},
  number={32},
  pages={3615--3642},
  year={2011},
  publisher={Wiley Online Library}
}

@article{tang2011fine,
  title={Fine tuning the energetics of excited-state intramolecular proton transfer (ESIPT): white light generation in a single ESIPT system},
  author={Tang, Kuo-Chun and Chang, Ming-Jen and Lin, Tsung-Yi and Pan, Hsiao-An and Fang, Tzu-Chien and Chen, Kew-Yu and Hung, Wen-Yi and Hsu, Yu-Hsiang and Chou, Pi-Tai},
  journal={Journal of the American Chemical Society},
  volume={133},
  number={44},
  pages={17738--17745},
  year={2011},
  publisher={ACS Publications}
}

@article{zhang2016control,
  title={Control of the reversibility of excited-state intramolecular proton transfer (ESIPT) reaction: host-polarity tuning white organic light emitting diode on a new thiazolo [5, 4-d] thiazole ESIPT system},
  author={Zhang, Zhiyun and Chen, Yi-An and Hung, Wen-Yi and Tang, Wei-Feng and Hsu, Yen-Hao and Chen, Chi-Lin and Meng, Fan-Yi and Chou, Pi-Tai},
  journal={Chemistry of Materials},
  volume={28},
  number={23},
  pages={8815--8824},
  year={2016},
  publisher={ACS Publications}
}

@article{mamada2017highly,
  title={Highly efficient thermally activated delayed fluorescence from an excited-state intramolecular proton transfer system},
  author={Mamada, Masashi and Inada, Ko and Komino, Takeshi and Potscavage Jr, William J and Nakanotani, Hajime and Adachi, Chihaya},
  journal={ACS Central Science},
  volume={3},
  number={7},
  pages={769--777},
  year={2017},
  publisher={ACS Publications}
}

@article{li2018dual,
  title={Dual-emissive 2-(2'-hydroxyphenyl) oxazoles for high performance organic electroluminescent devices: discovery of a new equilibrium of excited state intramolecular proton transfer with a reverse intersystem crossing process},
  author={Li, Bijin and Zhou, Linsen and Cheng, Hu and Huang, Quan and Lan, Jingbo and Zhou, Liang and You, Jingsong},
  journal={Chemical Science},
  volume={9},
  number={5},
  pages={1213--1220},
  year={2018},
  publisher={Royal Society of Chemistry}
}

@article{wu2018novo,
  title={De novo design of excited-state intramolecular proton transfer emitters via a thermally activated delayed fluorescence channel},
  author={Wu, Kailong and Zhang, Tao and Wang, Zian and Wang, Lian and Zhan, Lisi and Gong, Shaolong and Zhong, Cheng and Lu, Zheng-Hong and Zhang, Song and Yang, Chuluo},
  journal={Journal of the American Chemical Society},
  volume={140},
  number={28},
  pages={8877--8886},
  year={2018},
  publisher={ACS Publications}
}

@article{zhou2018unraveling,
  title={Unraveling the detailed mechanism of excited-state proton transfer},
  author={Zhou, Panwang and Han, Keli},
  journal={Accounts of Chemical Research},
  volume={51},
  number={7},
  pages={1681--1690},
  year={2018},
  publisher={ACS Publications}
}

@article{fang2009mapping,
  title={Mapping GFP structure evolution during proton transfer with femtosecond Raman spectroscopy},
  author={Fang, Chong and Frontiera, Renee R and Tran, Rosalie and Mathies, Richard A},
  journal={Nature},
  volume={462},
  number={7270},
  pages={200--204},
  year={2009},
  publisher={Nature Publishing Group UK London}
}

@article{han2013excited,
  title={Excited-state proton transfer of photoexcited pyranine in water observed by femtosecond stimulated Raman spectroscopy},
  author={Han, Fangyuan and Liu, Weimin and Fang, Chong},
  journal={Chemical Physics},
  volume={422},
  pages={204--219},
  year={2013},
  publisher={Elsevier}
}

@article{houari2014modeling,
  title={Modeling optical signatures and excited-state reactivities of substituted hydroxyphenylbenzoxazole (HBO) ESIPT dyes},
  author={Houari, Ym{\`e}ne and Charaf-Eddin, Azzam and Laurent, Ad{\`e}le D and Massue, Julien and Ziessel, Raymond and Ulrich, Gilles and Jacquemin, Denis},
  journal={Physical Chemistry Chemical Physics},
  volume={16},
  number={4},
  pages={1319--1321},
  year={2014},
  publisher={Royal Society of Chemistry}
}

@article{laurent2014dye,
  title={Dye chemistry with time-dependent density functional theory},
  author={Laurent, Adele D and Adamo, Carlo and Jacquemin, Denis},
  journal={Physical Chemistry Chemical Physics},
  volume={16},
  number={28},
  pages={14334--14356},
  year={2014},
  publisher={Royal Society of Chemistry}
}

@article{chrayteh2020dual,
  title={Dual fluorescence in strap ESIPT systems: a theoretical study},
  author={Chrayteh, Amara and Ewels, Chris and Jacquemin, Denis},
  journal={Physical Chemistry Chemical Physics},
  volume={22},
  number={2},
  pages={854--863},
  year={2020},
  publisher={Royal Society of Chemistry}
}

@article{aquino2005excited,
  title={Excited-state intramolecular proton transfer: a survey of TDDFT and RI-CC2 excited-state potential energy surfaces},
  author={Aquino, Adelia JA and Lischka, Hans and H{\"a}ttig, Christof},
  journal={The Journal of Physical Chemistry A},
  volume={109},
  number={14},
  pages={3201--3208},
  year={2005},
  publisher={ACS Publications}
}

@article{aquino2009ultrafast,
  title={Ultrafast excited-state proton transfer processes: energy surfaces and on-the-fly dynamics simulations},
  author={Aquino, Ad{\'e}lia JA and Plasser, Felix and Barbatti, Mario and Lischka, Hans},
  journal={Croatica Chemica Acta},
  volume={82},
  number={1},
  pages={105--114},
  year={2009},
  publisher={Hrvatsko kemijsko dru{\v{s}}tvo}
}

@article{sobolewski2009computational,
  title={Computational studies of the photophysics of neutral and zwitterionic amino acids in an aqueous environment: tyrosine-(H2O) 2 and tryptophan-(H2O) 2 clusters},
  author={Sobolewski, Andrzej L and Shemesh, Dorit and Domcke, Wolfgang},
  journal={The Journal of Physical Chemistry A},
  volume={113},
  number={3},
  pages={542--550},
  year={2009},
  publisher={ACS Publications}
}

@article{verite2019theoretical,
  title={A theoretical elucidation of the mechanism of tuneable fluorescence in a full-colour emissive ESIPT dye},
  author={V{\'e}rit{\'e}, Pauline M and H{\'e}d{\'e}, Simon and Jacquemin, Denis},
  journal={Physical Chemistry Chemical Physics},
  volume={21},
  number={31},
  pages={17400--17409},
  year={2019},
  publisher={Royal Society of Chemistry}
}

@article{chrayteh2020td,
  title={TD-DFT and CC2 insights into the dual-emissive behaviour of 2-(2'-hydroxyphenyl) oxazoles core and their derivatives},
  author={Chrayteh, Amara and Ewels, Chris P and Jacquemin, Denis},
  journal={Physical Chemistry Chemical Physics},
  volume={22},
  number={43},
  pages={25066--25074},
  year={2020},
  publisher={Royal Society of Chemistry}
}

@article{sobolewski1999ab,
  title={Ab initio potential-energy functions for excited state intramolecular proton transfer: a comparative study of o-hydroxybenzaldehyde, salicylic acid and 7-hydroxy-1-indanone},
  author={Sobolewski, Andrzej L and Domcke, Wolfgang},
  journal={Physical Chemistry Chemical Physics},
  volume={1},
  number={13},
  pages={3065--3072},
  year={1999},
  publisher={Royal Society of Chemistry}
}

@article{li2015casscf,
  title={A CASSCF/CASPT 2 insight into excited-state intramolecular proton transfer of four imidazole derivatives},
  author={Li, Yuanyuan and Wang, Li and Guo, Xugeng and Zhang, Jinglai},
  journal={Journal of Computational Chemistry},
  volume={36},
  number={32},
  pages={2374--2380},
  year={2015},
  publisher={Wiley Online Library}
}

@article{chang2023caspt2,
  title={CASPT2//CASSCF studies on mechanistic photophysics of 3-hydroxyflavone},
  author={Chang, Xue-Ping and Fan, Feng-Ran and Zhao, Geng and Ma, Xiantao and Zhang, Teng-Shuo and Xie, Bin-Bin},
  journal={Chemical Physics},
  volume={575},
  pages={112056},
  year={2023},
  publisher={Elsevier}
}

@article{ortiz2007electronic,
  title={Electronic and quantum dynamical insight into the ultrafast proton transfer of 1-hydroxy-2-acetonaphthone},
  author={Ortiz-S{\'a}nchez, Juan Manuel and Gelabert, Ricard and Moreno, Miquel and Lluch, Jos{\'e} M},
  journal={The Journal of Chemical Physics},
  volume={127},
  number={8},
  pages={084318},
  year={2007},
  publisher={AIP Publishing}
}

@article{perveaux2017fast,
  title={Fast and slow excited-state intramolecular proton transfer in 3-hydroxychromone: a two-state story?},
  author={Perveaux, Aurelie and Lorphelin, Maxime and Lasorne, Benjamin and Lauvergnat, David},
  journal={Physical Chemistry Chemical Physics},
  volume={19},
  number={9},
  pages={6579--6593},
  year={2017},
  publisher={Royal Society of Chemistry}
}

@article{anand2020excited,
  title={Excited-state intramolecular proton transfer driven by conical intersection in hydroxychromones},
  author={Anand, Neethu and Isukapalli, Sai Vamsi Krishna and Vennapusa, Sivaranjana Reddy},
  journal={Journal of Computational Chemistry},
  volume={41},
  number={11},
  pages={1068--1080},
  year={2020},
  publisher={Wiley Online Library}
}

@article{anand2020h,
  title={O--H vibrational motions promote sub-50 fs nonadiabatic dynamics in 3-hydroxypyran-4-one: interplay between internal conversion and ESIPT},
  author={Anand, Neethu and Nag, Probal and Kanaparthi, Ravi Kumar and Vennapusa, Sivaranjana Reddy},
  journal={Physical Chemistry Chemical Physics},
  volume={22},
  number={16},
  pages={8745--8756},
  year={2020},
  publisher={Royal Society of Chemistry}
}

@article{pijeau2017excited,
  title={Excited-state dynamics of 2-(2'-Hydroxyphenyl) benzothiazole: ultrafast proton transfer and internal conversion},
  author={Pijeau, Shiela and Foster, Donneille and Hohenstein, Edward G},
  journal={The Journal of Physical Chemistry A},
  volume={121},
  number={24},
  pages={4595--4605},
  year={2017},
  publisher={ACS Publications}
}

@article{pijeau2018effect,
  title={Effect of nonplanarity on excited-state proton transfer and internal conversion in salicylideneaniline},
  author={Pijeau, Shiela and Foster, Donneille and Hohenstein, Edward G},
  journal={The Journal of Physical Chemistry A},
  volume={122},
  number={25},
  pages={5555--5562},
  year={2018},
  publisher={ACS Publications}
}

@article{sporkel2013photodynamics,
  title={Photodynamics of schiff base salicylideneaniline: trajectory surface-hopping simulations},
  author={Sp\"{o}rkel, Lasse and Cui, Ganglong and Thiel, Walter},
  journal={The Journal of Physical Chemistry A},
  volume={117},
  number={22},
  pages={4574--4583},
  year={2013},
  publisher={ACS Publications}
}

@article{nag2021ultrafast,
  title={Ultrafast nonadiabatic excited-state intramolecular proton transfer in 3-hydroxychromone: A surface hopping approach},
  author={Nag, Probal and Anand, Neethu and Vennapusa, Sivaranjana Reddy},
  journal={The Journal of Chemical Physics},
  volume={155},
  number={9},
  pages={094301},
  year={2021},
  publisher={AIP Publishing}
}

@article{nag2022unraveling,
  title={Unraveling the sub-100 fs ESIPT in 5-hydroxychromone using surface hopping simulations},
  author={Nag, Probal and Vennapusa, Sivaranjana Reddy},
  journal={Journal of Photochemistry and Photobiology A: Chemistry},
  volume={427},
  pages={113767},
  year={2022},
  publisher={Elsevier}
}

@article{zhao2024excited,
  title={Excited-state dynamics of 3-hydroxychromone in gas phase},
  author={Zhao, Li and Geng, Xuehui and Wang, Jiangyue and Liu, Yuxuan and Yan, Wenhui and Xu, Zhijie and Chen, Junsheng},
  journal={Physical Chemistry Chemical Physics},
  volume={26},
  number={30},
  pages={20490--20499},
  year={2024},
  publisher={Royal Society of Chemistry}
}

@article{ash2011excited,
  title={Excited state intramolecular proton transfer in 3-hydroxychromone: a DFT-based computational study},
  author={Ash, Sankarlal and De, Sankar Prasad and Beg, Hasibul and Misra, Ajay},
  journal={Molecular Simulation},
  volume={37},
  number={11},
  pages={914--922},
  year={2011},
  publisher={Taylor \& Francis}
}

@article{estiu1999theoretical,
  title={A theoretical study of excited state proton transfer in 3-hydroxychromone and related molecules},
  author={Esti{\'u}, Guillermina and Rama, Jos{\'e} and Pereira, Alberto and Cachau, Ra{\'u}l E and Ventura, Oscar N},
  journal={Journal of Molecular Structure: THEOCHEM},
  volume={487},
  number={3},
  pages={221--230},
  year={1999},
  publisher={Elsevier}
}

@article{kenfack2012ab,
  title={Ab initio study of the solvent H-bonding effect on ESIPT reaction and electronic transitions of 3-hydroxychromone derivatives},
  author={Kenfack, Cyril A and Klymchenko, Andrey S and Duportail, Guy and Burger, Alain and M{\'e}ly, Yves},
  journal={Physical Chemistry Chemical Physics},
  volume={14},
  number={25},
  pages={8910--8918},
  year={2012},
  publisher={Royal Society of Chemistry}
}

@article{chevalier2013esipt,
  title={ESIPT and photodissociation of 3-hydroxychromone in solution: Photoinduced processes studied by static and time-resolved {UV}/{V}is, fluorescence, and IR spectroscopy},
  author={Chevalier, Katharina and Gru{\"u}n, Anneken and Stamm, Anke and Schmitt, Yvonne and Gerhards, Markus and Diller, Rolf},
  journal={The Journal of Physical Chemistry A},
  volume={117},
  number={44},
  pages={11233--11245},
  year={2013},
  publisher={ACS Publications}
}

@article{adamo1999toward,
  title={Toward reliable density functional methods without adjustable parameters: The PBE0 model},
  author={Adamo, Carlo and Barone, Vincenzo},
  journal={The Journal of Chemical Physics},
  volume={110},
  number={13},
  pages={6158--6170},
  year={1999},
  publisher={American Institute of Physics}
}

@article{dunning1989gaussian,
  title={Gaussian basis sets for use in correlated molecular calculations. I. The atoms boron through neon and hydrogen},
  author={Dunning Jr, Thom H},
  journal={The Journal of Chemical Physics},
  volume={90},
  number={2},
  pages={1007--1023},
  year={1989},
  publisher={American Institute of Physics}
}

@misc{g16,
author={M. J. Frisch and G. W. Trucks and H. B. Schlegel and G. E. Scuseria and M. A. Robb and J. R. Cheeseman and G. Scalmani and V. Barone and G. A. Petersson and H. Nakatsuji and X. Li and M. Caricato and A. V. Marenich and J. Bloino and B. G. Janesko and R. Gomperts and B. Mennucci and H. P. Hratchian and J. V. Ortiz and A. F. Izmaylov and J. L. Sonnenberg and D. Williams-Young and F. Ding and F. Lipparini and F. Egidi and J. Goings and B. Peng and A. Petrone and T. Henderson and D. Ranasinghe and V. G. Zakrzewski and J. Gao and N. Rega and G. Zheng and W. Liang and M. Hada and M. Ehara and K. Toyota and R. Fukuda and J. Hasegawa and M. Ishida and T. Nakajima and Y. Honda and O. Kitao and H. Nakai and T. Vreven and K. Throssell and Montgomery, {Jr.}, J. A. and J. E. Peralta and F. Ogliaro and M. J. Bearpark and J. J. Heyd and E. N. Brothers and K. N. Kudin and V. N. Staroverov and T. A. Keith and R. Kobayashi and J. Normand and K. Raghavachari and A. P. Rendell and J. C. Burant and S. S. Iyengar and J. Tomasi and M. Cossi and J. M. Millam and M. Klene and C. Adamo and R. Cammi and J. W. Ochterski and R. L. Martin and K. Morokuma and O. Farkas and J. B. Foresman and D. J. Fox},
title={Gaussian~16 {R}evision {A}.03},
year={2016},
note={Gaussian Inc. Wallingford CT}
}

@article{neese2025software,
  title={Software update: The ORCA program system—version 6.0},
  author={Neese, Frank},
  journal={Wiley Interdisciplinary Reviews: Computational Molecular Science},
  volume={15},
  number={2},
  pages={e70019},
  year={2025},
  publisher={Wiley Online Library}
}

@article{maeda2010updated,
  title={Updated branching plane for finding conical intersections without coupling derivative vectors},
  author={Maeda, Satoshi and Ohno, Koichi and Morokuma, Keiji},
  journal={Journal of Chemical Theory and Computation},
  volume={6},
  number={5},
  pages={1538--1545},
  year={2010},
  publisher={ACS Publications}
}

@article{mai2018nonadiabatic,
  title={Nonadiabatic dynamics: The SHARC approach},
  author={Mai, Sebastian and Marquetand, Philipp and Gonz{\'a}lez, Leticia},
  journal={Wiley Interdisciplinary Reviews: Computational Molecular Science},
  volume={8},
  number={6},
  pages={e1370},
  year={2018},
  publisher={Wiley Online Library}
}

@incollection{crespo2012spectrum,
  title={Spectrum simulation and decomposition with nuclear ensemble: formal derivation and application to benzene, furan and 2-phenylfuran},
  author={Crespo-Otero, Rachel and Barbatti, Mario},
  booktitle={Marco Antonio Chaer Nascimento: A Festschrift from Theoretical Chemistry Accounts},
  pages={89--102},
  year={2012},
  publisher={Springer}
}

@article{pitesa2024excitonic,
  title={Excitonic configuration interaction: Going beyond the Frenkel exciton model},
  author={Pitesa, Tomislav and Polonius, Severin and Gonz{\'a}lez, Leticia and Mai, Sebastian},
  journal={Journal of Chemical Theory and Computation},
  volume={20},
  number={13},
  pages={5609--5634},
  year={2024},
  publisher={ACS Publications}
}

@article{granucci2001direct,
  title={Direct semiclassical simulation of photochemical processes with semiempirical wave functions},
  author={Granucci, Giovanni and Persico, Maurizio and Toniolo, Alessandro},
  journal={The Journal of Chemical Physics},
  volume={114},
  number={24},
  pages={10608--10615},
  year={2001},
  publisher={American Institute of Physics}
}

@article{granucci2007critical,
  title={Critical appraisal of the fewest switches algorithm for surface hopping},
  author={Granucci, Giovanni and Persico, Maurizio},
  journal={The Journal of Chemical Physics},
  volume={126},
  number={13},
  pages={134114},
  year={2007},
  publisher={AIP Publishing}
}

@article{levine2006conical,
  title={Conical intersections and double excitations in time-dependent density functional theory},
  author={Levine, Benjamin G and Ko, Chaehyuk and Quenneville, Jason and Mart{\'i}nez, Todd J},
  journal={Molecular Physics},
  volume={104},
  number={5-7},
  pages={1039--1051},
  year={2006},
  publisher={Taylor \& Francis}
}

@article{McGibbon2015MDTraj,
    title = {MDTraj: A Modern Open Library for the Analysis of
    Molecular Dynamics Trajectories},
    author = {McGibbon, Robert T. and Beauchamp, Kyle A. and Harrigan,
    Matthew P. and Klein, Christoph and Swails, Jason M. and
    Hern{\'a}ndez, Carlos X.  and Schwantes, Christian R. and Wang,
    Lee-Ping and Lane, Thomas J. and Pande, Vijay S.},
    journal = {Biophysical Journal},
    volume = {109},
    number = {8},
    pages = {1528 -- 1532},
    year = {2015},
    doi = {10.1016/j.bpj.2015.08.015}
}

@article{pearson1901liii,
  title={LIII. On lines and planes of closest fit to systems of points in space},
  author={Pearson, Karl},
  journal={The London, Edinburgh, and Dublin philosophical magazine and journal of science},
  volume={2},
  number={11},
  pages={559--572},
  year={1901},
  publisher={Taylor \& Francis}
}

@misc{hastie2009elements,
  title={The Elements of Statistical Learning},
  author={Hastie, Trevor and Tibshirani, Robert and Friedman, Jerome and others},
  year={2009},
  publisher={Springer series in statistics New-York}
}

@article{fisher1936use,
  title={The use of multiple measurements in taxonomic problems},
  author={Fisher, Ronald A},
  journal={Annals of Eugenics},
  volume={7},
  number={2},
  pages={179--188},
  year={1936},
  publisher={Wiley Online Library}
}

@article{klymchenko20043,
  title={3-Hydroxychromone dyes exhibiting excited-state intramolecular proton transfer in water with efficient two-band fluorescence},
  author={Klymchenko, Andrey S and Demchenko, Alexander P},
  journal={New journal of chemistry},
  volume={28},
  number={6},
  pages={687--692},
  year={2004},
  publisher={Royal Society of Chemistry}
}

@article{giordano2023tuning,
  title={Tuning of environment-sensitive 3-hydroxychromone fluorophores based on strong donor substituents in positions 2 or 7},
  author={Giordano, Luciana and Shvadchak, Volodymyr V and Arrupe, Nicolas and Lockhart, Lisandro J Falomir and Sanchez, Veronica M and Jovin, Thomas M},
  journal={Dyes and Pigments},
  volume={218},
  pages={111479},
  year={2023},
  publisher={Elsevier}
}

@article{bersuker2013pseudo,
  title={Pseudo-Jahn--Teller Effect--A Two-State Paradigm in Formation, Deformation, and Transformation of Molecular Systems and Solids},
  author={Bersuker, Isaac B},
  journal={Chemical Reviews},
  volume={113},
  number={3},
  pages={1351--1390},
  year={2013},
  publisher={ACS Publications}
}

@article{lipparini2010variational,
  title={A variational formulation of the polarizable continuum model},
  author={Lipparini, Filippo and Scalmani, Giovanni and Mennucci, Benedetta and Canc{\`e}s, Eric and Caricato, Marco and Frisch, Michael J},
  journal={The Journal of chemical physics},
  volume={133},
  number={1},
  pages={014106},
  year={2010},
  publisher={AIP Publishing}
}

\end{document}